\documentclass[11pt]{article}
\usepackage{pstricks}
\usepackage{pstricks-add}
\usepackage{xcolor}
\usepackage{mathrsfs}
\usepackage{enumerate}    
\usepackage{type1cm}        
\usepackage{makeidx}         
\usepackage{graphicx}        
\usepackage{multicol}        
\usepackage[bottom]{footmisc}
\usepackage{newtxtext}       %
\usepackage{newtxmath}       

\newcommand{\RR}{\mbox{$\mathbb{R}$}}
\newcommand{\R}{\mathcal{R}}
\newcommand{\Ro}{\mathcal{R}_0}
\newcommand{\Ho}{\mathcal{H}_0}
\newcommand{\Vo}{\mathcal{V}_0}
\newcommand{\Rov}{\mathcal{J}_{0}}

\newcommand{\dfe}{\mbox{dfe}}

\newcommand{\tcr}{\textcolor{black}}

\newcommand{\tcm}{\textcolor{black}}
\newtheorem{theorem}{Theorem}[section]

\newtheorem{remark}{Remark}[section]
\newtheorem{proposition}{Proposition}[section]

\newcommand{\bpf}{{\sl \bf Proof :}\hspace{5mm}}
\begin{document}
\title{Application of Mathematical Epidemiology to crop \\ vector-borne diseases. The cassava mosaic virus disease case.}
%
\author{Michael Chapwanya$^1$ and Yves Dumont$^{1,2,3,4}$ \footnote{Corresponding author: yves.dumont@cirad.fr} \\
\small{$^1$University of Pretoria, Department of Mathematics and Applied Mathematics, Pretoria, South Africa}\\ 
\small{$^2$CIRAD, UMR AMAP, F-34398 Montpellier, France} \\
\small{$^3$AMAP, Univ Montpellier, CIRAD, CNRS, INRA, IRD, Montpellier, France} \\
\small{$^4$EPITAG, LIRIMA, France}}
\date{}
\maketitle

\abstract{In this chapter, an application of Mathematical Epidemiology to crop vector-borne diseases is presented to investigate the interactions between crops, vectors, and virus. The main illustrative example is the cassava mosaic disease (CMD). The CMD virus has two routes of infection: through vectors and also through infected crops. In the field, the main tool to control CMD spreading is roguing. The presented biological model is sufficiently generic and the same methodology can be adapted to other crops or crop vector-borne diseases. After an introduction where a brief history of crop diseases and useful information on Cassava and CMD is given, we develop and study a compartmental temporal model, taking into account the crop growth and the vector dynamics. A brief qualitative analysis of the model is provided, i.e., existence and uniqueness of a solution, existence of a disease-free equilibrium and existence of an endemic equilibrium. We also provide conditions for local (global) asymptotic stability and show that a Hopf Bifurcation may occur, for instance, when diseased plants are removed. Numerical simulations are provided to illustrate all possible behaviors. Finally, we discuss the theoretical and numerical outputs in terms of crop protection.}



\section{Introduction}\label{sec:intro}

\tcm{Agriculture began from around $9000$ BC independently in different places around the world. Since then, humans have developed their knowledge to produce food more efficiently and to increase crop yields, while promoting the development of the human community. However, this is possible as long as Mother Earth has the capacity to sustain all developments induced by humans. Looking ahead to the year 2050, approximately $9$ billion people are expected to live in places where 10 000 years ago, only a few hundred thousands were living. In essence, there is a challenge to provide food for people on (more and more) limited area of (arable) land. Even if crop yields have drastically increased during the last century, crops have to face new dangers in the form of diseases and pests.  For a number of years, pesticides were developed as a solution to combat the spread of pests and diseases. Unfortunately, we now know that most of these pesticides have caused a lot of damage (like cancer, lost in the biodiversity, etc). Consequently, there is need to develop sustainable approaches to maintain yields and reduce the use of chemical products as much as possible in order to protect the biodiversity, decrease the risk of cancer or other diseases, and, also, to protect our Earth for the future generations.}

While the first mathematical model on inoculation against smallpox was developed by Daniel Bernoulli around 1760, it is mostly acknowledged that Mathematical Epidemiology and the first compartmental models were initiated, not by mathematicians, but by public health physicians, like Sir R.A. Ross, W.H. Hamer, A.G. McKendrick, and W.O. Kermack between 1900 and 1935, \cite{Brauer2017} (see also \cite{Bacaer}). Since then, various epidemiological models have been developed and studied mathematically. However, while mathematical models for vector borne diseases in humans and animals are well documented in the literature, very little work has been done in modeling vector-borne diseases in plants. 

The aim of this chapter is to present an example of crop-vector-virus interaction, using cassava crop because it has become a very important crop for many people in Africa. After rice and maize, it is now the third-largest source of food and carbohydrates, \cite{Fauquet1990}. We will begin here in the introduction by giving a brief historical background of crop diseases to emphasize that crop diseases are as old as human diseases like malaria, and probably started with the neolithic revolution when humans moved from a lifestyle of hunting and gathering to one of agriculture and settlement. We then explain the importance of cassava and the impact of its main disease, the cassava mosaic disease, and summarize previous modeling works.

\subsection{Crop diseases in the past}
Crop diseases are known since the ancient time (see for instance \cite{Lucas1992} and also \cite{Borkar2017}). Plant diseases have been recorded in a few Egyptian writings, on sumerian clay tablets (1200 BC), and also in the Old testament and in the Talmud. Even in the far East, plant diseases were recognized, in particular, in India, Vedic literature (1500-500 BC), and also some two thousands old works are available in China. 
In ancient Greece, some diseases of trees, cereals and vegetables have been reported by Theophrastus (371-287 BC), also known as the `father of Botany'. He was student, collaborator and successor of Aristotle. Theophrastus and Pliny the Elder (23 -79 AD), became the references in Europe until the end of the Middle Age. However, between the 8th and 13th century, the Arab agricultural revolution occurred. Ibn al-Awwam (Seville, late 12th) wrote one of the most important medieval book on Agriculture and also about the symptoms of tree and vine diseases as well as (more or less realistic) methods to cure. Aztec, Inca and Mayas in the Americas also faced diseases on maize and potatoes, that will be imported by conquistadors, first, in Spain and in Portugal (in \tcm{1493}, for maize, when Christopher Colombus went back to Spain \cite{Goodman}; potatoes were only recorded in the literature in 1537 after the Spaniards discovered high Andes of South America \cite{Hawkes}),  before their spreading throughout Europe along the 16th -17th centuries. They became major crops in many countries in the "Old World". Particularly potatoes, leading to one of the most notable historical impacts of plant disease, the Irish potato famine of 1845, that was caused by late blight fungus, \textit{Phytophtora infestans}. Approximately one million people perished from starvation; around one million and a half more are believed to have left Ireland and immigrated to the United States. The French vine industry was devastated in 1878 by another fungal pathogen, the powdery mildew. This is the first time where disease management was able to find a solution and produce the first fungicide ever, the Bordeaux mixture (made with copper sulfate and lime mixture). However, for the vine industry, the real disaster came with the phylloxera, a sap sucking insects, that first devastated most of the vineyards in the continent before spreading to the rest of the world, except some places, like Chile. In fact, the number of diseases (and also pest attacks) have increased dramatically since the 19th century, with the increase in population mobility and also with the increase of trade exchanges, including the importation of alien plants/plant products containing diseases or insects (vectors of diseases). What is unusual with crop vector-borne diseases is that in general one vector (aphids for instance) is able to transmit several to hundreds of viruses on many crops. So that several crops can be impacted on at the same time by several virus causing more damages and eventually death. Since the discoveries of new lands, the importation of new crops, the increase of trades and population displacements, diseases and vectors have invaded or are invading areas where plants are fully susceptible, such that, in the recent decades, pests and diseases have become a major problem in many crops around the world, and, especially in Africa. 

\subsection{About cassava, \tcm{cassava mosaic disease,} and its vector}
This chapter focuses on cassava ({\it Manihot esculenta} Crantz) and the cassava mosaic virus. Cassava was domesticated in the southwestern Amazon of Brazil and Bolivia some 8,000-10,000 years ago (see \cite{Isendahl} for further details). It was introduced from the Americas onto Africa by Portuguese traders in the 16th century. Since then it has spread on the African continent, such that, after rice and corn, it has become a very important staple food for over 500 millions people in Africa alone. Cassava can grow in places where other crops cannot. It is tolerant to drought and can grow on soils with a low nutrient capacity. Cassava is a perennial crop. However the storage roots can be harvested from 6 to 24 months after planting, depending on the cultivar and the growing conditions. Moreover, the roots can be left in the ground without harvesting for a long period of time, making it a very useful crop as a security against famine. That is why cassava is considered as a food security crop. Its study and its protection against bio-aggressors are of main importance. The current average yields per hectare is around 12.5 tonnes, while the optimal yields can be as high as 80 tonnes under optimal conditions. This shows that there is potential for increase in production. More than 1 million of small-scale farmers around the world are producing cassava. See \cite{FAO2013} for further information.

Cassava mosaic geminiviruses (CMGs) are the causal agents of cassava mosaic disease (CMD), which is one of the most widespread and devastating diseases of cassava in Africa. For instance, in the 1990s a severe form of the CMD lead to yield losses of $80$ to $100$ percent in Uganda and Kenya (FAO, 2015). Currently, nine distinct CMG species (seven African and two Indian) have been found to infect cassava \cite{Patil2009}.

CMD is now prevalent in many parts of Africa \cite{Tresh2005}. The geminiviruses and their whitefly vector (\textit{Bemisia tabaci}) have been studied extensively and much attention has been given to possible control measures and their deployment since the 1990s. CMD was first described in Tanzania in 1894. Till then, CMD became prevalent in most of the countries of sub-Saharan Africa due to the continuous cultivation of susceptible cultivars and the indiscriminate use of diseased propagation material. As highlighted in \cite{holt} (an references therein), CMD is also propagated routinely from stem cuttings coming from infected plants, such that the virus can be maintained within the plant waiting for vectors to spread.
Early studies found that some varieties of cassava were less affected than others by CMD, so that resistance programs were launched in the 1930s-1940s. In the 1980s and 1990s, many projects aimed at controlling CMD were launched in Kenya, Ivory Coast and Uganda, such that CMD received more attention than any other crop disease and this also explains why CMD is always so prominent in the literature on cassava. However, whatever the research efforts done in the last thirty years, CMD is still a problem.

Cassava mosaic geminiviruses are transmitted in a persistent manner by whiteflies, \textit{B. tabaci} (see \cite{Macfayden2018} for a recent review). This was first showed in 1932 in the Democratic Republic of Congo. Further studies showed that whiteflies are the sole vectors. There are some vectors that can acquire the virus but are unable to transmit it. The minimum time between acquisition and becoming infected is around 7 hours. However, the transmission efficiency is very low ($0.15-1.7 \%$ for insect collected in the field \cite{Fargette1985}, to moderate ($4-13\%$) for laboratory-reared insects \cite{Maruthi2002}. The CMGs spreading is mainly related to the susceptibility/resistance of the cassava varieties, the phytosanitation measures and if CMGs are systemic within the infected plants. However, clearly, the mobility of the whiteflies is a key factor and in general higher incidence due to severe CMD is mainly due to a high level of infestation by \textit{B. tabaci}. Last but not least \textit{B. tabaci} also causes direct damages to cassava by feeding on phloem or deposition of honeydew, which acts as a substrate for sooty moulds (a black, non-parasitic, superficial growth of fungi on  plant surfaces), that reduces both respiration and photosynthesis \cite{Nelson2008}. Surprisingly, as highlighted in \cite{Macfayden2018}, the research effort on the ecology of these insects was very little compared to all work done on the virus and resistant cassava cultivars. However, \textit{B. tabaci} is a polyphagous herbivore that can potentially use a wide range of different host plants (more than 500) in cassava production landscape so that, in general, intercropping cassava with other plants (maize, coffee, sweet potato, bean, etc) may result in a lower \textit{B. tabaci} population abundance. However, the mechanisms resulting in the decay are not necessarily related to host-preference. Coming back to the biology of \textit{B. tabaci}, published information suggests that its development period from egg to adult emergence is between $19$ and $29$ days, and the species goes through four nymphal instars before entering a pupal phase. However, in this paper, we will not enter in such details since we mainly focus on \textit{B. tabaci} as vector of the CMD. Note also that \textit{B. tabaci} is able to transmit more than $150$ viruses, such that it has become a very serious pest, not only for cassava. In particular, another virus that is really detrimental to cassava, also transmitted by \textit{B. tabaci}, is the cassava brown streak virus (CBSV) \cite{Legg2015}.

The symptoms of CMD are very well known. Common symptoms include misshapen leaves, chlorosis, mottling and mosaic. Less severe symptoms are ill-defined mosaic patterns, green mosaic with slight or absence of leaf distortion. Thus, in general, symptoms range from barely perceptible mosaic to stunting and general decline of plants. The more severe the symptoms, the lower the yield. In general young plants are more severely infected than old ones. Last but not least, the symptoms are even more severe in the case of co-infection by two cassava strains, like ACMV (African Cassava Mosaic Virus) and EACMV (Eastern-Africa Cassava Mosaic Virus). There is a synergistic interaction between the two viruses.

Since the early CMD epidemics, phytosanitary strategies have been developed. Of course, these control strategies should be sustainable, and preferably with no use of pesticides. Following \cite{Tresh2005}, three approaches to decrease losses due to a virus disease have been proposed: 
\begin{enumerate}[$(i)$]
\item decrease the proportion of infected plants; 
\item delay infection to such a late stage of crop growth so that losses become unimportant; 
\item decrease the severity of damage sustained after infection in diverse ways. 
\end{enumerate}
These objectives can be achieved using phytosanitation, disease-resistant varieties, good cultural practices, vector-control and mild-strain protection. All previous control measures, combined or not, have been more or less studied. In this work, we will particularly explore roguing (the removal of diseased/infected plants), that is a very well known means of virus disease control. It is now clear that whiteflies are not easy to control by insecticides, such that vector control seems to be ineffective. However, we will also explore other control approaches that are used by small-scale farmers for other crops against whiteflies that are not necessarily used to protect cassava.


\subsection{On the usefulness of Mathematical Modeling \tcm{of cassava mosaic disease}}
The previous summary on Cassava and Cassava Mosaic Virus shows that despite years of study, field investigations and field experiments, the risk of severe epidemics like in the 1990's cannot be excluded. We believe it is not enough
to launch new research programs based only on field experiments/observations and laboratory experiments. We need to consider additional tools such as mathematical modeling and computer simulations. Mathematical modeling offers a cheaper, convenient and powerful tool to explore the Cassava-CMD system and go further in order to better understand the possible dynamics that can occur and to possibly point out new ways worth investigating. 
Some mathematical models on CMD have already been developed, see for instance \cite{Fargette1994,holt,Jeger2004}, but they were mostly studied using numerical simulations, without providing the qualitative analysis of the proposed mathematical models. Of course, here we distinguish compartmental models from statistical models, which are, in general, based on data sets. The paper from Holt et al. \cite{holt} provides many interesting results. However their models are mainly based on the fact that \textit{B. tabaci} population is limited by cassava individuals, which, in an African context, is not really true since in the introduction, we emphasize that the host range for \textit{B. tabaci} is large and that in general cassava is inter-cropped with other plants. Thus the vector-population is modeled using a logistic-like equation in the susceptible compartment, assuming some kind of skip-oviposition in the birth-rate functional, but this is not clearly explained. They also assumed constant replanting and additional mortality rate for infected plants. Both crop and vector interact using mass-action principle. Since, it has been highlighted that \textit{B. tabaci} population size can be very large, the use of a simple mass-action function could be questionable: in fact it is not the case here since the vector population is limited by the number of cassava individuals (which is a restriction, as explained above).
However, the reader has to be aware that this is the way the modeling works: you consider and/or give modeling assumptions to build your model(s), such that for the same biological problem you can derive many different models, see for example the discussion in \cite{jeger2018}. Using their models \cite{holt}, Holt and collaborators obtained various results, mainly based on numerical simulations. Their study showed a range of dynamic behavior including cycles of infection. In their conclusions, clearly selection the cuttings and roguing are the main tools to lower the incidence of the disease. In \cite{Zhang2000} and more recently in \cite{Hebert2016}, authors included vector aggregation (and also dispersal in \cite{Hebert2016}) in their models, through nonlinear acquisition and inoculation rates. These previous works highlight the wide diversity and complexity of plant-vector-virus modeling. \tcm{However, among all these models and also many others in crop epidemiology, plants are always considered as \tcr{mature} individuals, ignoring their growth and the effect of vectors on them before and during outbreaks \cite{chapwanya_dumont}. }






The structure of the chapter is as follows: in section \ref{sec:mathmodel}, we build the mathematical model related to Cassava-CMVD interactions taking into account plant biomass growth. In section \ref{sec:qualitative}, a qualitative analysis is provided. In section \ref{sec:perma}, a specific study about the permanence of the disease is done. Finally, in section \ref{sec:numerical}, a sensitivity analysis followed by numerical simulations are given to illustrate the theoretical results. \tcm{In particular, we discuss the usefulness of roguing and eventually other control options that could be tested with our approach.}

\section{The mathematical model}\label{sec:mathmodel}
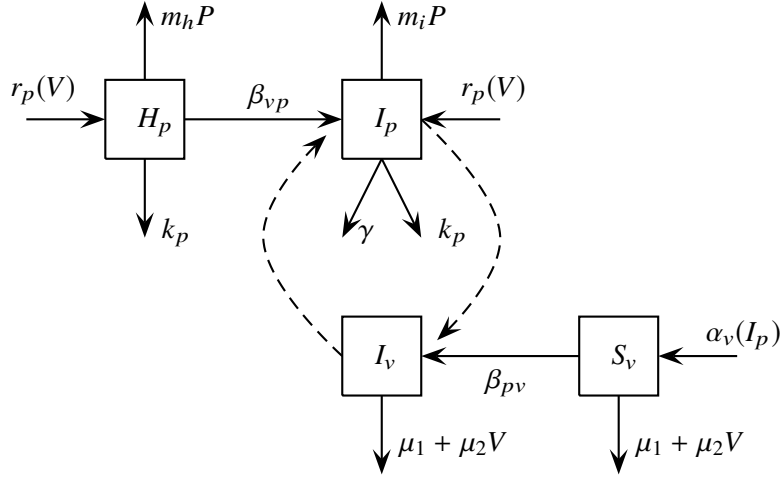
\begin{figure}[ht]
\begin{center}
\begin{pspicture}(0,0)(12,6.5)\psset{unit=1.05cm}
\psset{arrowsize=0.23}
\psline{-}(2,4)(2,5)(3,5)(3,4)(2,4)\put(2.4,4.4){$H_p$} \psline{->}(1,4.5)(2,4.5) \psline{->}(2.5,5)(2.5,6)\put(2.7,5.7){$m_hP$} \put(0.8,4.8){$r_{p }(V)$}\psline{->}(2.5,4)(2.5,3)\put(2.7,3){$k_p$}
\psline{-}(5,4)(5,5)(6,5)(6,4)(5,4)\put(5.4,4.4){$I_p$} \psline{->}(5.5,5)(5.5,6) \put(5.7,5.7){$m_iP$} \psline{<-}(6,4.5)(7,4.5)\put(6.5,4.8){$r_{p }(V)$}\psline{->}(5.5,4)(5,3)\put(5.2,3){$\gamma$}\psline{->}(5.5,4)(6,3)\put(6.2,3){$k_p$}
\psline{-}(5,1)(5,2)(6,2)(6,1)(5,1)\put(5.4,1.4){$I_v$} \psline{->}(5.5,1)(5.5,0)\put(5.7,0.3){$\mu_1+\mu_2V$} 
\psline{-}(8,1)(8,2)(9,2)(9,1)(8,1)\put(8.4,1.4){$S_v$}\psline{->}(8.5,1)(8.5,0)\put(8.7,0.3){$\mu_1+\mu_2V$} \psline{->}(10,1.5)(9,1.5)\put(9.6,1.7){{$\alpha_v(I_p)$}}
\psline{->}(3,4.5)(5,4.5) \put(3.8,4.7){$\beta_{vp}$}
\psline{<-}(6,1.5)(8,1.5) \put(6.8,1.1){$\beta_{pv}$} 
\pscurve[linestyle=dashed,dotsep=0.1pt]{<-}(4.8,4.3)(4,3)(5,1.5) 
\pscurve[linestyle=dashed,dotsep=0.1pt]{->}(6,4.5)(7,3)(6.2,1.7) 
\end{pspicture}
\caption{Schematic diagram of the model. }
\label{figure1}
\end{center}
\end{figure}

The model is formulated by considering two compartments: the total {plant} biomass $P$, and the vector population {density} $V$. The total biomass is further subdivided into two disjoint epidemiological states: the susceptible healthy biomass $H_p$, and the infectious biomass $I_p$, so that $P=H_p+I_p$. Similarly, the vector population is divided into two disjoint analogous classes, the susceptible vectors $S_v$, and the infectious vector $I_v$ so that $V=S_v+I_v$.
Based on the above knowledge, we propose the epidemiological structure represented by the compartmental diagram presented in Fig. \ref{figure1}, page \pageref{figure1}, which leads to the following system of equations
\begin{equation}\left\{
\begin{aligned}\label{model1}
\dfrac{dH_{p}}{dt}&=r_{p }(V)H_{p}-{m_h}PH_{p}-\beta_{vp}H_{p}I_{v}{-k_pH_p},\\
\dfrac{dI_{p}}{dt}&=r_{p }(V)I_p+\beta_{vp}H_{p}I_{v}-{m_i}PI_p-\gamma I_{p}{-k_pI_p},\\
\dfrac{dS_{v}}{dt}&={\alpha_{v}(I_p)}V-\left(\mu_{1}+\mu_{2}V\right)S_{v}-\beta_{pv}S_{v}I_{p} ,\\
\dfrac{dI_{v}}{dt}&=\beta_{pv}S_{v}I_{p} - \left( \mu_{1}+\mu_{2}V\right)I_{v},
\end{aligned}
\right.
\end{equation}
with initial conditions
\[
H_p(0)=H_p^0\geq0,\quad I_p(0)=I_p^0\geq0, \quad S_v(0)=S_v^0\geq0, \quad I_v(0)=I_v^0\geq0,
\]
where all the parameters are positive. Let us now describe the formulation of model \eqref{model1} in detail.

Following \cite{Lebon2014} (see also \cite{Paine2012}), the crop biomass growth is assumed to be a nonlinear process, influenced by maintenance (respiration, etc) and intra-specific competition, while the biomass production (through photosynthesis, reproduction, etc) is assumed to be a linear process of the biomass. Thus, the total biomass $P$ follows a logistic-like equation
\[
\dfrac{dP}{dt}=(r_{{p}}(V)-k_p)P-m_h(P)P,
\]
where $r_{{p}}(V)$ is the biomass production rate, $k_p$ is the harvest rate, and the function $m_h(P)$ represents the losses of biomass due to respiration, transpiration and maintenance. It is a positive increasing function of the biomass \cite{Lebon2014}. According to \cite{Keating1982}, where the authors fit cassava (dry) biomass growth with a logistic equation, we set $m_h(P)=m_hP$.
{Our crop growth model is very "simple" compared to previous process-based models (PBM) (like for instance GUMCAS \cite{Mathews1994}, GUMCAS-DSSAT \cite{Gabriel2014}, and SIMANIHOT \cite{Tironi2017}), but sufficient to explore the dynamics of the disease, while keeping the model mathematically tractable. Last but not least, the use of the previous PBM needs to estimate all the physiological parameters not only on healthy plants but also on diseased plants. These data are in general never available.}

\tcm{The function $r_{p }(V)$ is assumed to be a decreasing and bounded function of $V$, {i.e., $r_{p }'(V)\leq 0$}. This makes sense since the increase in the vector population may negatively impact the production of new biomass.}

According to whiteflies biology, and like in \cite{chapwanya_dumont}, we assume that the vector population follows a logistic growth with $\alpha_v-\mu_1$ being the net growth rate, and $\mu_2$ being the death rate due to the density effect. Contrary to \cite{holt}, we don't rely the dynamics on the presence/absence of cassava, since in the African context, many other cultivated crops can host whiteflies. In fact, we follow \cite{Jeger2004}, where the authors revisited the Holt Cassava Model by considering a logistic equation to model the whiteflies dynamic. However, we take into account that whiteflies have a negative effect on Cassava growth rate, such that the growth rate function is decreasing with increasing $V$, \tcm{through a parameter $\phi$, related to sap sucking}. We also assume that all vector newborns are susceptible and, according to the mean lifespan of whiteflies, we assume that all infected vectors stay infected for the rest of their lives (the impact of the viruliferous period can be very important within a spatio-temporal model \cite{chapwanya_dumont}).

The presence of the virus in the host and in the vector populations has several effects. We model only two known effects that were not necessarily developed in previous models:
\begin{enumerate}[$(i)$]
    \item First, infected plants grow less than susceptible plants: the maintenance of the infected biomass, $m_i$, is larger than $m_h$, the maintenance parameter of healthy plant, due to plant defences mechanisms. 
    \item Second, the vector growth rate, $\alpha_v-\mu_1$, can also be impacted positively by the presence of diseased plants (see \cite{Colvin2006}), so that we consider $\alpha_v(I_p)-\mu_1$.
\end{enumerate}

Infection between the plants and vectors is  modelled by the mass-action principle. The parameter $\beta_{vp}$ represents the contact rate between infectious vectors and susceptible plants. Similarly, $\beta_{pv}$ represents the contact rate between infectious plants and susceptible vectors. The assumption is that the environment is uniform, homogeneous and randomly mixed. In the model we also introduce the roguing parameter $\gamma$, i.e., the rate at which infected biomass is removed. In addition, $k_p$ is the rate at which the cassava is harvested for their tuberous roots. The model assumes both the susceptible and infected biomass can be harvested.
\tcm{A summary (and values) of all parameters is provided in Table \ref{tabsensi}, page \pageref{tabsensi}. Detailed explanations of the given range of values is provided at the beginning of section \ref{sec:numerical}, page \pageref{sec:numerical}.}


Following \eqref{model1}, when CMV is circulating, $P$ and $V$ satisfy the following equations
\begin{equation}\label{eqP}
\frac{dP}{dt}=(r_{p }(V){-k_p})P-({m_hH_p+m_iI_p})P-\gamma I_p,
\end{equation}
and
\begin{equation}\label{eqV}
\frac{dV}{dt}={\alpha_{v}(I_p)} V-(\mu_1+\mu_2V)V.
\end{equation}

\begin{table}[h!]
  \centering
  \begin{tabular}{|c|c|c|c|c|}
\hline
Parameters & Name & Range & Unit & Source \\ 
\hline
$r$       & biomass maximal growth rate & $[0.01,0.06]$ & day$^{-1}$ & estimated \tcr{from} \cite{Keating1982} 
\\
\hline
$\phi$     & impact of sap sucking per vector & $[0,0.02]$ & vector$^{-1}$ & \tcr{assumed} \\ \hline
$m_h$  & maintenance rate per unit of biomass& $[0.003,0.01]$ & biomass$^{-1}$day$^{-1}$ &estimated \tcr{from} \cite{Keating1982}  \\ \hline
$m_i\geq m_h$ & maintenance rate per unit of biomass   & $[0.01,0.1]$ & biomass$^{-1}$day$^{-1}$ & \tcr{assumed} \\
& when the biomass is infected & & & \\ \hline
$\beta_{vp}$  & infection rate & $[0.002,0.032]$ & vector$^{-1}$day$^{-1}$ & \cite{holt} \\ \hline
$\gamma$ & roguing rate & $[0,1]$ & day$^{-1}$ & \tcr{assumed}  \\ \hline
$\alpha$ &  whitefly rate of increase & $[0.1,1]$ & day$^{-1}$ & estimated \tcr{from} \cite{Jeger2004} \\ \hline
$\mu_1$  & average death rate & $[0.05,0.2]$& day$^{-1}$ &  estimated \tcr{from} \cite{Jeger2004} \\ \hline
$\mu_2$ & average density death rate & $[0.0001,0.01]$ & vector$^{-1}$day$^{-1}$ & estimated \tcr{from} \cite{Jeger2004}\\ \hline
$\beta_{pv}$& acquisition rate & $[0.01,0.1]$ & biomass$^{-1}$day$^{-1}$ & \tcr{assumed}  \\ \hline
$k_p$ & mean harvest rate& $[0.0025,0.0072]$ & day$^{-1}$ & estimated \tcr{from} \cite{holt}  \\ \hline
$\delta$ & boost parameter related to $I_p$ & $[0,1]$ & - & \tcr{assumed}  \\ \hline
\end{tabular}\\
  \caption{Range of values for the parameters of system~(\ref{model1}).}\label{tabsensi}
\end{table}

\section{Qualitative analysis}\label{sec:qualitative}
In this section we take $r_{{p}}(V)={r}/{(1+\phi V)}$, which satisfies the assumption above for $\phi,r>0$ {with $r_{p }(V)\leq r_{{p}}(0)=r$}. 
{In addition, a possible choice for $\alpha_v$ is  $\alpha_v(I_p)=\alpha\left(1+\dfrac{\delta I_p}{1+I_p}\right)$ which satisfies $\alpha_v'(I_p)>0$ and $\alpha \leq \alpha_v(I_p)\leq\alpha(1+\delta)$ for all $I_p$.}
We start with the well-posedness result which we state as follows.
\begin{theorem}
Assume {$\alpha_v(0)>\mu_1$ and $r_{p }(0)>k_p$}.
The system defines a dynamical system on the biologically feasible region
\[ \Omega=\left\{ (H_p,I_p,S_v,I_v)\in\RR^4_+: H_p+I_p\leq P^{\dag}, \; S_v+I_v\leq V^\dag  \right\} \]
where  $V^\dag=\dfrac{{\alpha(1+\delta)}-\mu_1}{\mu_2}$ and $P^{\dag}=\dfrac{{r_{{p}}(0)}-k_p}{m_h}$.
\end{theorem}

\bpf
We want to show that for any non negative initial data, the system possesses for all $t\geq0$, a unique solution which lies in the region $\Omega$. We will provide the proof in two steps.

First we want to show that $\Omega$ is a positively invariant set. In particular, we want to show that no trajectory leaves $\Omega$ by crossing one of its faces, \cite{Busenberg1993}. On the contrary, let us assume there exist $t_1>0$ such that $H_p(t_1)=0$, $H_p'(t_1)<0$ with $H_p(t)>0,\; I_p(t)>0,\; S_v(t)>0,\; I_v(t)>0$ for all $t\in(0,t_1)$. The first equation gives
\begin{equation*}
\frac{d H_p(t_1)}{d t} = 0,
\end{equation*}
which is a contradiction. Therefore $H_p(t)\geq0$ for all $t\geq0$. Similarly for $t_2>0$ with $I_p(t_2)=0$, $I_p'(t_2)<0$ with $H_p(t)>0,\; S_v(t)>0,\; I_v(t)>0$ for all $t\in(0,t_2)$, we have
\begin{equation*}
\frac{d I_p(t_2)}{d t} = \beta_{vp}H_pI_v>0,
\end{equation*}
and $t_3>0$ with $S_v(t_3)=0$, $S_v'(t_3)<0$ with $H_p(t)>0,\; I_p(t)>0,\; I_v(t)>0$ for all $t\in(0,t_3)$, we have
\begin{equation*}
 \frac{d S_v(t_3)}{d t} = \alpha_v(I_p)I_v>0, 
\end{equation*}
and $t_4>0$ with $I_v(t_4)=0$, $I_v'(t_4)<0$ with $H_p(t)>0,\; I_p(t)>0,\; S_v(t)>0$ for all $t\in(0,t_4)$, we have
\begin{equation*}
 \frac{d I_v(t_4)}{d t} = \beta_{pv}S_vI_p>0.
\end{equation*}
Therefore, in all cases the solution remains in $\Omega$ for any initial data in $\Omega$. 

In the second step, we use the a prior estimate derived below together with the fact that the right-hand side of the system is a locally Lipschitz function.

From equation \eqref{eqP}, assuming $I_p\geq0$ we have the inequality
\[
\frac{dP}{dt}\leq(r_{p }(V){-k_p})-P({m_hH_p+m_iI_p}) \leq ({r_{p }(0)}-k_p)-{m_h}P^2,
\]
so that $\displaystyle{\lim_{t\rightarrow \infty}\sup P(t)\leq \frac{{r_{p }(0)}-k_p}{{m_h}}}$, {where we have used the assumption that $r_{p }(V)>k_p$}. Similarly, from equation \eqref{eqV} we have  $\displaystyle{\lim_{t\rightarrow \infty}\sup V(t)\leq \frac{{\alpha}(1+\delta)-\mu_1}{\mu_2}}$ provided {$\alpha_v(0)>\mu_1$}. That is all the solutions are uniformly bounded. Combining the above two steps we conclude that \eqref{model1} defines a dynamical system on $\Omega$.

 \begin{flushright}
     $\Box$
 \end{flushright}


\subsection{Computation of $\mathcal{R}_{0}$}
{In this section we use the next generation matrix approach described in \cite{vandendriessche2002} (see also \cite{diekmann2009}), to compute the basic reproduction number $\R_0$, of the system. In our case, this is defined as the number of secondary infections that an infectious crop or vector could produce when introduced in a healthy population.}

The system \eqref{model1} has a disease-free equilibrium $E_{DFE}=(P^*,0,V^*,0)$. We consider only the equations where the infection progress, namely
\begin{equation}\label{eq:ngm1}
\dfrac{dX}{dt}=\mathcal{F}(X)-\mathcal{V}(X),
\end{equation}
where, in this case $X=(I_p,I_v)^t$, and
\[
\mathcal{F}(X)=\left(\begin{array}{c}
r_{p }(V)I_p+\beta_{vp}H_{p}I_{v},\\
\beta_{pv}S_{v}I_{p}
\end{array}\right)\quad\mbox{ and }\quad\mathcal{V}(X)=\left(\begin{array}{c}
\left({m_i}P+\gamma{+k_p} \right)I_p\\
\left(\mu_{1}+\mu_{2}V\right)I_{v}
\end{array}\right).
\]
{Equation \eqref{eq:ngm1} distinguishes the rate of appearance of new infections $\mathcal{F}(X)$, from the transfer into or out of the population by any other means, $\mathcal{V}(X)$.} Then
\[
J_{F}(X)=\left(\begin{array}{cc}
r_{p }(V) & -r^2_v(V)\phi/r{I_p}+\beta_{vp}H_{p}\\
\beta_{pv}S_{v} & 0
\end{array}\right)
\]
and
\[
J_{V}(X)=\left(\begin{array}{cc}
{m_i}I_p+{m_i}P+\gamma{+k_p} & 0\\
0 & \mu_{1}+\mu_{2}V+\mu_{2}I_v
\end{array}\right).
\]
We set
\[
F=\left(\begin{array}{cc}
{r_{p }(V^*)} & \beta_{vp}P^{*}\\
\beta_{pv}V^{*} & 0
\end{array}\right)
\]
and
\[
V=\left(\begin{array}{cc}
{m_i}P^*+\gamma{+k_p} & 0\\
0 & \mu_{1}+\mu_{2}V^{*}
\end{array}\right)=\left(\begin{array}{cc}
({m_i}-m_h)P^*+\gamma+r_{p }(V^*) & 0\\
0 & {\alpha_v(0)}
\end{array}\right),
\]
{where $P^*=\dfrac{r_{p }(V^*)-k_p}{{m_h}}$} and $V^*=\dfrac{\alpha_v(0)-\mu_1}{\mu_2}$. Then the next generation matrix $K$ is
\[
FV^{-1}=\left(\begin{array}{cc}
{r_{p }(V^*)} & \beta_{vp}P^{*}\\
\beta_{pv}V^{*} & 0
\end{array}\right)\left(\begin{array}{cc}
\dfrac{1}{({m_i}-m_h)P^*+\gamma+r_{p }(V^*)} & 0\\
0 & {\dfrac{1}{\alpha_v(0)}}
\end{array}\right),
\]
that is
\[
FV^{-1}=
\left(\begin{array}{cc}
\dfrac{{r_{p }(V^*)}}{({m_i}-m_h)P^*+\gamma+r_{p }(V^*)} & \dfrac{\beta_{vp}P^{*}}{\alpha_v(0)}\\
\dfrac{\beta_{pv}V^{*}}{({m_i}-m_h)P^*+\gamma+r_{p }(V^*)} & 0
\end{array}\right).
\]
From which we deduce that

\begin{equation}
\mathcal{R}_0=\frac{1}{2}\left(\mathcal{V}_0+\sqrt{\mathcal{V}_0^2+4\mathcal{H}_0}\right),
\label{R0}
\end{equation}
 where
    \[
    \mathcal{V}_0=\frac{r_{p }(V^*)}{(m_i-m_h)P^*+\gamma+r_{p }(V^*)},
    \]
    represent a vertical transmission (transmission through the production of new plant biomass), and
    \[
    \mathcal{H}_0=\frac{\beta_{pv}\beta_{vp}V^*P^*}{\alpha_v(0)[(m_i-m_h)P^*+\gamma+r_{p }(V^*)]}=\frac{\beta_{pv}\beta_{vp}V^*P^*}{\alpha_v(0)[m_iP^*+\gamma+k_p]}
    \]
    represent an horizontal transmission (through vectors)

\begin{remark}\label{remark0}
Note also that when $(m_i-m_h)P^*+\gamma=0$, that is when there is no roguing, i.e., $\gamma=0$, and $m_i=m_h$, we have $\mathcal{V}_0=1$, such that we always have $\mathcal{R}_0>1$.
\end{remark}

\begin{remark}
\tcm{Instead of the complex expression given in (\ref{R0}), for the basic reproduction ratio, it would be biologically more natural to consider $\mathcal{V}_0+\sqrt{\mathcal{H}_0}$, with both ways of transmission: the one-phase infectious process, from plant to plant, and the two-phase infectious process, from vector to plant and from plant to vector. Indeed, $\mathcal{V}_0+\sqrt{\mathcal{H}_0}<1$ implies $\mathcal{R}_0<1$, and $\mathcal{R}_0>1$ implies $\mathcal{V}_0+\sqrt{\mathcal{H}_0}>1$.}
\end{remark}
Finally, following \cite{vandendriessche2002}, we deduce the following result
\begin{theorem}
When $\mathcal{R}_0<1$ the DFE is locally asymptotically stable, while unstable when $\mathcal{R}_0>1$.
\end{theorem}


\subsection{Trivial equilibria and their stabilities/unstabilities}

Several equilibria exist for model \eqref{model1}. In this section we focus on the following equilibria,
\[
 E_{+}=(0,0,V^{*},0), \quad\quad E_{++}=(P^\dag,0,0,0),
\]
with $P^\dag=\dfrac{r_{p }(0)-k_p}{m_h}>P^*{=\dfrac{r_{p }(V^*)-k_p}{m_h}}$; the disease-free equilibrium
\[
E_{DFE}=\left(P^{*},0,V^{*},0\right).
\]
Computing the Jacobian of system \eqref{model1} leads to
\[ J=
\left(
\begin{array}{cccc}
f(H_p,I_p,S_v,I_v)  & -{m_h}H_p  & -\dfrac{r\phi H_p}{(1+\phi V)^2} & -\dfrac{r\phi H_p}{(1+\phi V)^2}-\beta_{vp}H_p  \\
 \beta_{vp}I_v-{m_i}I_p & g(H_p,I_p,S_v,I_v)  & -\dfrac{r\phi I_p}{(1+\phi V)^2} & -\dfrac{r\phi I_p}{(1+\phi V)^2}+\beta_{vp}H_p \\
0  & {\alpha_v'(I_p)V-\beta_{pv}S_v}  & {h(H_p,I_p,S_v,I_v)}  & {\alpha_v(I_p)-\mu_2 S_v} \\
0  & \beta_{pv}S_v   & \beta_{pv}I_p-\mu_2 I_v  & -\mu_2(I_v+V)-\mu_1
\end{array}
\right),
\]
where $f(H_p,I_p,S_v,I_v)=r_{p }(V)-{m_h}(P+H_p)-\beta_{vp}I_v{-k_p}$, $g(H_p,I_p,S_v,I_v)=r_{p }(V)-{m_i}(P+I_p)-\gamma{-k_p}$ {and $h(H_p,I_p,S_v,I_v)=\alpha_v(I_p)-(\mu_1+\mu_2V)-\beta_{pv}I_p-\mu_2 S_v$}. Thus at $E_+$ we have 
\begin{equation}
J_{E_+}=
\left(
\begin{array}{cccc}
r_{p }(V^*){-k_p}  & 0  & 0 & 0  \\
0  &  r_{p }(V^*)-\gamma{-k_p} & 0 & 0 \\
0  & {\alpha_v'(0)V^*-\beta_{pv}V^*}  & {-\alpha_v(0)-\mu_2 V^*}  & {\alpha_v(0)-\mu_2 V^*} \\
0  & \beta_{pv}V^*   & 0  & -\mu_2V^*-\mu_1
\end{array}
\right).
\end{equation}
Expanding in terms of the first column, it is clear that at least one of the eigenvalues is strictly positive, which implies that $E_{+}$ is unstable. At $E_{++}$ we have
\begin{equation}
J_{E_{++}}=
\left(
\begin{array}{cccc}
r_{p }(0)-k_p  & -{m_h}P^\dag  & -r\phi P^\dag & -P^\dag(r\phi+1)  \\
0  &  -\gamma{-k_p} & 0 & \beta_{vp}P^\dag \\
0  & 0  & \alpha_v(0)-\mu_1  & {\alpha_v(0)} \\
0  & 0   & 0  & -\mu_1
\end{array}
\right).
\end{equation}
$J_{E_{++}}$ being an upper triangular matrix, the  eigenvalues are the diagonal terms. Since  $\alpha_v(0)-\mu_1$ is strictly positive, $E_{++}$ is unstable.

At the disease-free-equilibrium, we have
\[ J_{\text{DFE}}=
\left(
\begin{array}{cccc}
-(r_{p }(V^*)-k_p)  & -{m_h}P^*  & -\dfrac{\phi P^*{r_{p }^*}^2}{r} & -\dfrac{\phi P^*{r_{p }^*}^2}{r}-\beta_{vp}P^*  \\
0  &  -({m_i}-m_h)P^*-\gamma & 0 & \beta_{vp}P^* \\
0  & {\alpha_v'(0)-\beta_{pv}V^*}  & -\mu_2 V^*  & {\alpha_v(0)-\mu_2 V^*} \\
0  & \beta_{pv}V^*   & 0  & -\mu_2V^*-\mu_1
\end{array}
\right).
\]
Considering $|J_{\text{DFE}}-\lambda I|$, we expand in terms of the first column and this gives $\lambda_1=-(r_{p }(V^*){-k_p}) <0$ {since $r_{p }(V^*)>k_p$}. We then expand the remaining matrix in terms of the second column and $\lambda_2=-\mu_2V^*$. The remaining two eigenvalues are eigenvalues of the matrix 
\[
\left(
\begin{array}{cc}
-({m_i}-m_h)P^*-\gamma &  \beta_{vp}P^* \\
\beta_{pv}V^*   &  -\mu_1-\mu_2 V^*
\end{array}
\right),
\]
or
\[
\left(
\begin{array}{cc}
-\left(({m_i}-m_h)P^*+\gamma\right)  &  \beta_{vp}P^* \\
\beta_{pv}V^*   &  {-\alpha_v(0)}
\end{array}
\right).
\]
after applying the definition of $P^*$ and $V^*$ to the diagonal entries of the reduced matrix. Clearly the trace of this matrix is negative and the requirement for positive determinant simplifies to \\ $\left(({m_i}-m_h)P^*+\gamma\right){\alpha_v(0)}-\beta_{pv} \beta_{vp}P^*V^*>0$, that is if \begin{equation}\label{R0v}
\mathcal{J}_{0}=\dfrac{\beta_{pv} \beta_{vp}P^*V^*}{\left(({m_i}-m_h)P^*+\gamma\right){\alpha_v(0)}}<1.
\end{equation}

We summarize the previous results as follows

\begin{proposition}
Consider model \eqref{model1}:
\begin{itemize}
\item The equilibria $E_+$ and $E_{++}$ are always unstable. 
\item Assuming $({m_i}-m_h)P^*+\gamma>0$. The disease-free equilibrium, $DFE$, is locally asymptotically stable when $\mathcal{J}_{0}<1$, and unstable otherwise.
\end{itemize}
\end{proposition}

\begin{remark}
When $({m_i}-m_h)P^*+\gamma=0$, according to the previous section, and remark \ref{remark0}, page \pageref{remark0}, we have $\mathcal{R}_0>1$, such that DFE is always unstable.
\end{remark}

\begin{remark}\label{remark1}
A remark is necessary here with regards to the basic reproduction numbers defined in \eqref{R0} and \eqref{R0v}. They are mathematically equivalent: as indicated in \cite{cushing} it will depend on the choice of $\mathcal{F}$. It is however important to have a biologically relevant decomposition. Using the next generation matrix approach, if we choose $r_{p }(V)I_p$ as part of $\mathcal{V}$, then $\Ro^2=\mathcal{J}_{0}$. 
However a direct computation shows that since $\Ro=1$ so is $\mathcal{J}_{0}$. Indeed, following \eqref{R0}, $\Ro=1$ is equivalent to 
$$
\sqrt{\Vo+4\Ho}=2-\Vo,
$$
or equivalently $\Ho=1-\Vo$, i.e.,
$$
\dfrac{\beta_{pv}\beta_{vp}P^*V^*}{[({m_i}-m_h)P^*+\gamma+r_{p }(V^*)]{\alpha_v(0)}}=1-\frac{r_{p }(V^*)}{({m_i}-m_h)P^*+\gamma+r_{p }(V^*)},
$$
or
$$
\dfrac{\beta_{pv}\beta_{vp}P^*V^*}{{\alpha_v(0)}}=({m_i}-m_h)P^*+\gamma,
$$
that is exactly equivalent to have $\mathcal{J}_{0}=1$.
More precisely, using the previous result, we can show the following relationships
$$
\Ro \leq \mathcal{J}_{0}\leq 1,\qquad \mbox{and}\qquad  1 \leq \mathcal{J}_{0} \leq \Ro.
$$
\end{remark}


\subsection{Global stability of the DFE}
Next we improve on the previous result and show that the disease free equilibrium is globally asymptotically stable (GAS) when $\mathcal{R}_{0}\leq1$. Following the work of \cite{Castillo_Chavez1998}, we rewrite our system by splitting the uninfected compartments $x=(H_{p},S_{v})$ from the infected compartments $y=(I_{p},I_{v})$. That is
\[
\left\{ \begin{array}{c}
\dfrac{dx}{dt}=f(x,y),\\
\dfrac{dy}{dt}=g(x,y),
\end{array}\right.
\]
such that $g(x,0)=0$. Let us consider the system
\begin{equation}
\dfrac{dx}{dt}=f(x,0)=\left(\begin{array}{c}
\dfrac{rH_{p}}{1+\phi S_v}-{m_h}H_{p}^2\\
{\alpha_{v}(0)}S_{v}-\left(\mu_{1}+\mu_{2}S_{v}\right)S_{v}
\label{subsystem1}
\end{array}\right).
\end{equation}
The second equation is the well known logistic equation, for which we know that $V^*$ is GAS. Then, we deduce that $P^*$ is GAS for the first equation. Altogether the DFE $x^*=(P^*,V^*)$ is GAS for system \eqref{subsystem1}. Now we consider
\[
g(x,y)=\left(\begin{array}{c}
r_{p }(V)I_p+\beta_{vp}H_{p}I_{v}-\left({m_i}P+\gamma {+k_p}\right)I_{p}\\
\beta_{pv}S_{v}I_{p}-\left(\mu_{1}+\mu_{2}V\right)I_{v}
\end{array}\right)
\]
such that
\[
J_{g}(x,y)=\left(\begin{array}{cc}
r_{p }(V)-\left({m_i}P+{m_i}I_p+\gamma{+k_p}\right) & -\dfrac{r\phi I_p}{(1+\phi V)^2}+\beta_{vp}H_{p}\\
\beta_{pv}S_{v} & -\left(\mu_{1}+\mu_{2}V+\mu_{2}I_v\right)
\end{array}\right),
\]
and
\[
J_{g}(\dfe,0)=\left(\begin{array}{cc}
r_{p }(V^*)-\left({m_i}P^*+\gamma {+k_p}\right) & \beta_{vp}P^*\\
\beta_{pv}V^{*} & -\left(\mu_{1}+\mu_{2}V^{*}\right)
\end{array}\right).
\]
Then
\[
g(x,y)=\left(\begin{array}{cc}
-\left({(m_i-m_h)P^*+\gamma}\right) & \beta_{vp}P^*\\
\beta_{pv}V^{*} & {-\alpha_v(0)}
\end{array}\right)y
-
\left(\begin{array}{c}
\beta_{vp}\left(P^*-H_{p}\right)\\
\beta_{pv}(V^{*}-S_{v})
\end{array}\right).
\]
Now we have to show that $\hat{g}(x,y)\geq0$. This is true if $S_{v}\leq V^{*}$
(always verified), and if $H_{p}\leq P^*$ (always verified). 

In addition, 
\[
A=\left(\begin{array}{cc}
-\left({(m_i-m_h)P^*+\gamma}\right) & \beta_{vp}P^*\\
\beta_{pv}V^{*} & {-\alpha_v(0)}
\end{array}\right)
\]
is a Metzler Matrix. Thus using the result from \cite{Castillo_Chavez1998}, we have the following result.
\begin{theorem}\label{thm3}
Assume $\gamma+(m_i-m_h)P^*>0$. The disease-free equilibrium, (DFE) is globally asymptotically stable if {$\mathcal{J}_{0}\leq1$ and unstable for $\mathcal{J}_{0}>1$}.
\end{theorem}
In the next section we investigate the permanence of system \eqref{model1}, page \pageref{model1}, and, potentially, the existence and stability of endemic equilibria.

\section{Permanence of system \eqref{model1} -  Endemic equilibrium}\label{sec:perma}
Studying the existence (and the stability/instability) of endemic equilibria for system \eqref{model1} is not an easy task when $\alpha_v(I_p)\equiv \alpha$. However, in Annexe A, page \pageref{annexeA}, we show that that two types of endemic equilibria may exist (under conditions): a full endemic equilibrium (when the whole biomass is infected, i.e. $H_p=0$), and an endemic equilibrium (when healthy and infected biomasses co-exist). 

Since $\alpha_v(I_p)\geq \alpha_v(0)$, for all $I_p\geq 0$, we would like to show that a direct study is not always necessary and that it it is possible to consider the following sub-system of \eqref{model1} to derive interesting facts about the dynamics of system (\ref{model1}):
\begin{equation}
\left\{
\begin{array}{ll}
\label{model1alpha0}
\dfrac{dH_{p}}{dt}&=r_{p }(V)H_{p}-{m_h}H_{p}P-\beta_{vp}H_{p}I_{v}{-k_pH_p},\\
\dfrac{dI_{p}}{dt}&=r_{p }(V)I_p+\beta_{vp}H_{p}I_{v}-{m_i}PI_p-\gamma I_{p}{-k_pI_p},\\
\dfrac{dS_{v}}{dt}&=\alpha_v(0) V-\left(\mu_{1}+\mu_{2}V\right)S_{v}-\beta_{pv}S_{v}I_{p} ,\\
\dfrac{dI_{v}}{dt}&=\beta_{pv}S_{v}I_{p} -\left( \mu_{1}+\mu_{2}V\right)I_{v},
\end{array}
\right.
\end{equation}
Using the previous results, it is straightforward to show that system \eqref{model1alpha0} admits a unique non negative solution, $\underline{x}$. Let $x$ be the solution of \eqref{model1}. Then, assuming that $x(0)=\underline{x}(0)$, and using the comparison Theorem (see, for instance, Theorem 3.2, page 209, in \cite{Waltman}), we can deduce that $\underline{x}$ is a lower solution of system \eqref{model1}, that is \tcr{$x(t)\geq \underline{x}(t)$, for all $t>0$}. 

Thus, any sufficient conditions to show permanence of \eqref{model1alpha0}, i.e. $\underline{x(t)}>0$, will ensure permanence of \eqref{model1}, $x(t)>0$ for all $t>0$.

Assuming the dynamics of the vector to be very fast compared to the dynamics of the biomass, we assume quasi-steady state approximation such that $V=V^*$. Then, system \eqref{model1alpha0} reduces to
\begin{equation}\left\{\begin{aligned}\label{model2}
\dfrac{dH_{p}}{dt}&=r^*H_{p}-m_hPH_{p}-\beta_{vp}H_{p}I_{v}{-k_pH_p},\\
\dfrac{dI_{p}}{dt}&=r^*I_p+\beta_{vp}H_{p}I_{v}-m_iPI_p-\gamma I_{p}{-k_pI_p},\\
\dfrac{dI_{v}}{dt}&=\beta_{pv}(V^*-I_v)I_{p} -\alpha_vI_{v},
\end{aligned}\right.\end{equation}
where $r^*=r/(1+\phi V^*)$. For the equilibria we need to solve
\begin{equation}\begin{aligned}\label{model4}
r^*H_{p}^*-m_hP^*H_{p}^*-\beta_{vp}H_{p}^*I_{v}^*{-k_pH_p^*}&=0,\\
r^*I_p+\beta_{vp}H_{p}^*I_{v}^*-m_iP^*I_p^*-\gamma I_{p}^*{-k_pI_p^*}&=0,\\
\beta_{pv}(V^*-I_v^*)I_{p}^* -\alpha_vI_{v}^*&=0.
\end{aligned}\end{equation}
{Notice that the definition of $P^*$ here is different from earlier results.} Thus looking for equilibria in equation \eqref{model4} leads to
\begin{equation}
\left(r^*{-k_p}-m_hH_p-m_iI_p\right)P^*=\gamma I_{p},
\end{equation}
{that is}
\begin{equation}
\left(r^*{-k_p}-m_hP^*-(m_i-m_h)I_p^*\right)P^*=\gamma I_{p}^*,
\end{equation}
so that
\begin{equation}
\dfrac{r^*{-k_p}-m_hP^*}{(m_i-m_h)P^*+\gamma}P^*=I_{p}^*,
\end{equation}
assuming either $\gamma>0$ or $m_i>m_h$.
Thus, if $I_{p}=0$ then $P_{\dfe}=\dfrac{r^*{-k_p}}{m_h}$, and thus $I_{v}=0$
and $H_{p,\dfe}=P_{\dfe}$. 

Now assume that $0\leq\gamma{+k_p}<r^*$, and $H_{p}=0$, then from the
second equation and third equation we deduce,
\[
P^{\#}=I_p^{\#}=\dfrac{r^*-\gamma{-k_p}}{m_i},\qquad I_{v}^{\#}=\dfrac{\beta_{pv}P^{\#}}{\alpha_{v}+\beta_{pv}P^{\#}}V^{*}.
\]
The basic reproduction number for the reduced system \eqref{model2} becomes
\[
\mathcal{R}_{0}=\frac{1}{2}\left(\frac{r^*}{(m_i-m_h){P_{\dfe}}+\gamma+r^*}+\sqrt{\left(\frac{r^*}{(m_i-m_h){P_{\dfe}}+\gamma+r^*}\right)^2+4\dfrac{\beta_{pv}\beta_{vp}P^* V^*}{\left((m_i-m_h){P_{\dfe}}+\gamma+r^*\right)\alpha_v} }\right).
\]
Assume now that $H_{p}>0$, Then, using the first equation, we have
\[
\dfrac{r^*{-k_p}-m_hP^*}{\beta_{vp}}=I_{v}.
\]
Replacing $I_{v}$ in the third equation leads to 
\[
\beta_{pv}\left(V^{*}-I_{v}\right)I_{p}-\alpha_{v}I_{v}=\beta_{pv}\left(V^{*}-\dfrac{r^*{-k_p}-m_hP^*}{\beta_{vp}}\right)I_{p}-\alpha_{v}\dfrac{r^*{-k_p}-m_hP^*}{\beta_{vp}}=0.
\]
Replacing $I_{p}$ by $\dfrac{r^*{-k_p}-m_hP^*}{(m_i-m_h)P+\gamma}P$
leads to
\[
\beta_{pv}\left(V^{*}-\dfrac{r^*{-k_p}-m_hP^*}{\beta_{vp}}\right)\dfrac{r^*{-k_p}-m_hP^*}{(m_i-m_h)P^*+\gamma}P^*=\alpha_{v}\dfrac{r^*{-k_p}-mP^*}{\beta_{vp}}.
\]
Assuming that $\left(r^*{-k_p}-mP^*\right)\neq0$, we can simplify, such
that
\[
\beta_{pv}\left( \beta_{vp}V^{*}-r^*{+k_p}+m_hP^* \right)P^* = 
\alpha_{v}\left( (m_i-m_h)P^*+\gamma) \right),
\]
that is
\[
\beta_{pv}m_h{P^*}^{2}-\beta_{pv}\left(r^*{-k_p}-\beta_{vp}V^{*}-\alpha_{v}(m_i-m_h)\right)P^*-\gamma\alpha_{v}=0,
\]
which is a quadratic equation in $P$. The discriminant of this equation is 
\[
\Delta=\left(\beta_{pv}(r^*{-k_p}-\beta_{vp}V^{*})-\alpha_{v}(m_i-m_h)\right)^{2}+4\gamma\alpha_{v}\beta_{pv}m_h>0,
\]
such that there exists only one non-negative solution
\[
P^{*}=\dfrac{1}{2\beta_{pv}m_h}\left(\beta_{pv}\left(r^*{-k_p}-\beta_{vp}V^{*}-\alpha_{v}(m_i-m_h)\right)+\sqrt{\Delta}\right)
\]
Then we can deduce 
\[
I_{p}^*=\dfrac{r^*{-k_p}-m_hP^*}{(m_i-m_h)P^*+\gamma}P^*,\qquad H_{p}=P^{*}\left(1-\dfrac{r^*{-k_p}-m_hP^*}{(m_i-m_h)P^*+\gamma}\right)
\]
and
\[
I_{v}=\dfrac{r^*{-k_p}-m_hP^{*}}{\beta_{vp}}.
\]
We summarize the previous results as follows

\begin{proposition}

System \eqref{model2} always admits the following three (non zero) equilibria:
\begin{itemize}
\item The disease free Equilibrium $DFE=(P_{\dfe},0,0)$, whatever $\gamma\geq0$.
\item When $0\leq\gamma+{k_p}<r^*$, a Full Disease Equilibrium, $FDE=(0,P^{\#},I_{v}^{\#})$.
\item When $\gamma>0$, an endemic equilibrium $EE=(H_p^{*},I_{p}^{*},I_{v}^{*})$.
\end{itemize}
\end{proposition}
One way, to investigate the local (asymptotic) stability, is to compute the Jacobian
\begin{align*}
J(X)=&\left(\begin{array}{ccc}
r^*-k_p-m_hI_p-2m_hH_{p}-\beta_{vp}I_{v} & -m_hH_{p} & -\beta_{vp}H_{p}\\
\beta_{vp}I_{v}-m_iI_{p} & r^*-m_iP-m_iI_{p}-\gamma{-k_p} & \beta_{vp}H_{p}\\
0 & \beta_{pv}\left(V^{*}-I_{v}\right) & -\beta_{pv}I_{p}-\alpha_{v}
\end{array}\right) \\ =&\left(\begin{array}{ccc}
r^*-k_p-m_hI_p-2m_hH_{p}-\beta_{vp}I_{v} & -m_hH_{p} & -\beta_{vp}H_{p}\\
{r^*-k_p-m_hP-m_iI_{p}} & r^*-m_iP-m_iI_{p}-\gamma{-k_p} & \beta_{vp}H_{p}\\
0 & \beta_{pv}\left(V^{*}-I_{v}\right) & -\beta_{pv}I_{p}-\alpha_{v}
\end{array}\right).
\end{align*}
{The stability of the disease-free equilibrium follows from the definition of $\mathcal{R}_0$.} 
{However, in the particular case of $FDE=(0,P^{\#},I_{v}^{\#})$, we obtain the following Jacobian Matrix
\begin{align*}
J(FDE)=&\left(\begin{array}{ccc}
r^*-k_p-m_hP^{\#}-\beta_{vp}V^{\#} & 0 & 0\\
r^*{-k_p}-{(m_i+m_h)}P^{\#} & r^*-2m_iP^{\#}-\gamma{-k_p} & 0\\
0 & \beta_{pv}\left(V^{*}-V^{\#}\right) & -\left(\beta_{pv}P^{\#}+\alpha_{v}\right)
\end{array}\right),
\end{align*}
that is
\begin{align*}
J(FDE)=&\left(\begin{array}{ccc}
r^*-k_p-m_h\dfrac{r^*-\gamma{-k_p}}{m_i}-\beta_{vp}V^{\#} & 0 & 0\\
\gamma-m_hP^\# & -r^*+\gamma{+k_p} & 0\\
0 & \beta_{pv}\left(V^{*}-V^{\#}\right) & -\beta_{pv}P^{\#}-\alpha_{v}
\end{array}\right),
\end{align*}
or
\begin{align*}
J(FDE)=&\left(\begin{array}{ccc}
\left(r^*-k_p\right)\left(1-\dfrac{m_h}{m_i}\right)+\gamma\dfrac{m_h}{m_i}-\beta_{vp}V^{\#} & 0 & 0\\
\gamma-m_hP^\# & -r^*+\gamma{+k_p} & 0\\
0 & \beta_{pv}\left(V^{*}-V^{\#}\right) & -\beta_{pv}P^{\#}-\alpha_{v}
\end{array}\right),
\end{align*}
Clearly, the eigenvalues are the terms in the diagonal. They are negative if $\gamma+k_p<r^*$ (this is the condition to have existence of FDE), and $\gamma+ P^*\left(m_i-m_h\right)<\dfrac{m_i}{m_h}\beta_{vp} V^{\#}$. After manipulation, and using (\ref{R0v}), the last condition is equivalent to $\mathcal{J}_{0}>\dfrac{P^*}{P^{\#}}\left(1+\dfrac{\beta_{vp}P^{\#}}{\alpha_v}\right)>1$. In other words, when $\mathcal{J}_{0}$ is sufficiently large, the whole crop can become infected.}

Now we consider $EE=(H_{p}^{*},I_{p}^{*},I_{v}^{*})$. {We obtain
\begin{align*}
J(EE)=&\left(\begin{array}{ccc}
-m_hH_{p}^* & -m_hH_{p}^* & -\beta_{vp}H_{p}^*\\
{r^*-k_p-m_hP^*-m_iI_{p}} & r^*-m_iP^*-m_iI_{p}-\gamma{-k_p} & \beta_{vp}H_{p}^*\\
0 & \beta_{pv}\left(V^{*}-I_{v}^*\right) & -\beta_{pv}I_{p}^*-\alpha_{v}
\end{array}\right).
\end{align*}}

In that case, we know that $\gamma{+k_p}>r^*$,
such that the characteristic polynomial becomes 
\begin{equation}
p(x)=x^{3}+a_{2}x^{2}+a_{1}x+a_{0},
\label{poly}
\end{equation}
where, clearly,
\[
a_{2}=-\text{trace}(J(X^{*}))=m_hH_p^*+m_iP^{*}+m_iI_p^*-r^*+\gamma+k_p+\beta_{pv}I_{p}^{*}+\alpha_{v}>0.
\]
The coefficient $a_1$ is given by
\begin{align*}
a_{1}&=\sum_i{i<j}\left|\begin{array}{cc}
a_{ii} & a_{ij}\\
a_{ji} & a_{jj}
\end{array}\right|\\ 
&=\left|\begin{array}{cc}
-m_hH_{p}^* & -m_hH_{p}^*\\
{r^*-k_p-m_hP^*-m_iI_{p}} & {r^*-m_iP^*-m_iI_{p}-\gamma-k_p}
\end{array}\right|+\left|\begin{array}{cc}
-m_hH_{p} & -\beta_{vp}H_{p}\\
0 & -\beta_{pv}I_{p}-\alpha_{v}
\end{array}\right|\\&+
\left|\begin{array}{cc}
{r^*-m_iP^*-m_iI_{p}-\gamma-k_p} & \beta_{vp}H_{p}\\
\beta_{pv}\left(V^{*}-I_{v}\right) & -\beta_{pv}I_{p}-\alpha_{v}
\end{array}\right|.
\end{align*}
It is clear that
\[
\left|\begin{array}{cc}
-m_hH_{p}^{*} & -m_hH_{p}^{*}\\
{r^*-k_p-m_hP^*-m_iI_p^*} & {r^*-m_iP^*-m_iI_{p}-\gamma-k_p}
\end{array}\right|=mH_{p}^{*}((m_i-m_h)P^*+\gamma)>0,
\]
\[
\left|\begin{array}{cc}
-m_hH_{p} & -\beta_{vp}\\
0 & -\beta_{pv}I_{p}-\alpha_{v}
\end{array}\right|=m_hH_{p}^{*}\left(\beta_{pv}I_{p}+\alpha_{v}\right)>0,
\]
and
\begin{align*}
\left|\begin{array}{cc}
{r^*-m_iP^*-m_iI_{p}-\gamma-k_p} & \beta_{vp}H_{p}^{*}\\
\beta_{pv}\left(V^{*}-I_{v}\right) & -\beta_{pv}I_{p}-\alpha_{v}
\end{array}\right|&\\ =\left({m_iP^*+m_iI_{p}+\gamma+k_p-r^*}\right)&\left(\beta_{pv}I_{p}^{*}+\alpha_{v}\right)-\beta_{pv}\beta_{vp}H_{p}^{*}\left(V^{*}-I_{v}^{*}\right).
\end{align*}
However, from \eqref{model4}, page \pageref{model4} we have
\[
\beta_{pv}\beta_{vp}H_{p}V^{*}=\left(m_iP^{*}+\gamma-r^* -k_p\right)\left(\alpha_{v}+\beta_{pv}I_{p}^{*}\right)
\]
such that 
\[
\left|\begin{array}{cc}
{r^*-m_iP^*-m_iI_{p}-\gamma-k_p} & \beta_{vp}H_{p}^{*}\\
\beta_{pv}\left(V^{*}-I_{v}\right) & -\beta_{pv}I_{p}-\alpha_{v}
\end{array}\right|
= m_iI_p^*(\beta_{pv}I_p^*+\alpha_v)+\beta_{pv}\beta_{vp}H_{p}^{*}I_{v}^{*} >0.
\]
In fact 
\[
a_{1}=m_hH_{p}^{*}((m_i-m_h)P^*+\gamma) +(m_hH_{p}^{*}+m_iI_p^*)(\beta_{pv}I_p^*+\alpha_v)+\beta_{pv}\beta_{vp}H_{p}^{*}I_{v}^{*}>0.
\]
Thus we deduce that $a_{1}>0$. Finally
\begin{align*}
a_{0}&=-\left|\begin{array}{ccc}
-m_hH_{p} & -m_hH_{p} & -\beta_{vp}H_{p}\\
{r^*-k_p-m_hP^*-m_iI_p^*} & {r^*-m_iP^*-m_iI_{p}-\gamma -k_p} & \beta_{vp}H_{p}\\
0 & \beta_{pv}\left(V^{*}-I_{v}\right) & -\beta_{pv}I_{p}-\alpha_{v}
\end{array}\right| \\
&=m_hH_{p}^{*}\left|\begin{array}{cc}
{r^*-m_iP^*-m_iI_{p}-\gamma -k_p} & \beta_{vp}H_{p}\\
\beta_{pv}\left(V^{*}-I_{v}\right) & -\beta_{pv}I_{p}-\alpha_{v}
\end{array}\right|\\ &+
\left({r^*-k_p-m_hP^*-m_iI_p^*}\right)\left|\begin{array}{cc}
\beta_{pv}\left(V^{*}-I_{v}\right) & -\beta_{pv}I_{p}-\alpha_{v}\\
-mH_{p} & -\beta_{vp}H_{p}
\end{array}\right|.
\end{align*}
We have, as given above,
\begin{align*}
m_hH_{p}^{*}\left|\begin{array}{cc}
{r^*-m_iP^*-m_iI_{p}-\gamma -k_p}& \beta_{vp}H_{p}\\
\beta_{pv}\left(V^{*}-I_{v}\right) & -\beta_{pv}I_{p}-\alpha_{v}
\end{array}\right|\\ =m_hH_{p}^{*}\left(
m_iI_{p}^{*}\left(\beta_{pv}I_{p}^{*}+\alpha_{v}\right)+\beta_{vp}H_{p}^{*}\beta_{pv}I_{v}^{*}
\right)>0,
\end{align*}
and
\begin{align*}
\left({r^*-k_p-m_hP^*-m_iI_p^*}\right)&\left|\begin{array}{cc}
\beta_{pv}\left(V^{*}-I_{v}\right) & -\beta_{pv}I_{p}-\alpha_{v}\\
-m_hH_{p} & -\beta_{vp}H_{p}
\end{array}\right|\\ =-\left(m_hP^{*}+m_iI_{p}^{*}-r^*{+k_p}\right)&\left(\beta_{pv}\beta_{vp}H_{p}\left(V^{*}-I_{v}\right)+m_hH_{P}^{*}\left(\beta_{pv}I_{p}+\alpha_{v}\right)\right),
\end{align*}
with
\begin{align*}
m_hH_{p}^{*}\left(m_iI_{p}^{*}\left(\beta_{pv}I_{p}^{*}+\alpha_{v}\right)\right)-\left(m_hP^{*}+m_iI_{p}^{*}-r^*{+k_p}\right)m_hH_{P}^{*}\left(\beta_{pv}I_{p}+\alpha_{v}\right)\\=-\left(m_hP^{*}-r^*+k_p\right)m_hH_{P}^{*}\left(\beta_{pv}I_{p}+\alpha_{v}\right)
\end{align*}
such that 
\begin{align*}
a_{0}=m_hH_{p}^{*}\beta_{vp}H_{p}^{*}\beta_{pv}I_{v}^{*}-\left(m_hP^{*}-r^*{+k_p}\right)m_hH_{P}^{*}\left(\beta_{pv}I_{p}+\alpha_{v}\right)\\-\left(m_hP^{*}+m_iI_{p}^{*}-r^*{+k_p}\right)\beta_{pv}\beta_{vp}H_{p}\left(V^{*}-I_{v}\right)
\end{align*}
\begin{align*}
a_{0}=\beta_{pv}\beta_{vp}H_{p}I_{v}\left({m_hH_p^*+m_hP^{*}+m_iI_{p}^{*}-r^*+k_p}\right)-\left(m_hP^{*}-r^*{+k_p}\right)m_hH_{P}^{*}\left(\beta_{pv}I_{p}+\alpha_{v}\right)\\-\left(m_hP^{*}+m_iI_{p}^{*}-r^*{+k_p}\right)\beta_{pv}\beta_{vp}H_{p}^*V^{*}
\end{align*}
However, using the fact that
\[
r^*{-k_p}-m_hP^{*}=\beta_{vp}I_{v}^{*},
\]
and
\[
{H_{p}^{*}\left(\beta_{vp}I_{v}-m_iI_{p}\right)=\left(m_iI_{p}+\left[\gamma+k_p-r^*\right]\right)I_{p}>0},
\]
we have
\begin{align*}
a_{0}=\beta_{pv}\beta_{vp}H_{p}^*I_{v}^*\left({m_hH_p^*+m_hP^{*}+m_iI_{p}^{*}-r^*+k_p} \right)+\beta_{vp}I_{v}m_hH_{p}^{*}\left(\beta_{pv}I_{p}+\alpha_{v}\right)\\+\beta_{pv}\beta_{vp}H_{p}^{*}V^{*}\left(\beta_{vp}I_{v}-m_iI_{p}^{*}\right),
\end{align*}
i.e.,
\begin{align*}
a_{0}=\beta_{pv}\beta_{vp}H_{p}I_{v}\left({m_hH_p^*+m_hP^{*}+m_iI_{p}^{*}-r^*+k_p}\right)+\beta_{vp}I_{v}m_hH_{P}^{*}\left(\beta_{pv}I_{p}+\alpha_{v}\right)\\+\beta_{pv}\beta_{vp}V^{*}\left(m_iI_{p}+\gamma-r^*{+k_p}\right)I_{p}>0
\end{align*}
Thus the first assumption of Rough-Hurwitz is verified: $a_{i}>0$.

{We now consider $\gamma$, the roguing parameter, as a bifurcation parameter. Let us define}
\begin{align*}
\Delta (\gamma)=&a_{1}a_{2}-a_{0}\\=&[m_hH_{p}^{*}((m_i-m_h)P^*+\gamma) +(m_hH_{p}^{*}+m_iI_p^*) \beta_{pv}I_p^*+\alpha_v)+\beta_{pv}\beta_{vp}H_{p}^{*}I_{v}^{*}]\\ \times&[m_hH_p^*+m_iP^{*}+m_iI_p^*-r^*+\gamma+k_p+\beta_{pv}I_{p}^{*}+\alpha_{v}] \\ - &\beta_{pv}\beta_{vp}H_{p}^*I_{v}^*\left({m_hH_p^*+m_hP^{*}+m_iI_{p}^{*}-r^*+k_p} \right)+\beta_{vp}I_{v}m_hH_{p}^{*}\left(\beta_{pv}I_{p}+\alpha_{v}\right)\\+& \beta_{pv}\beta_{vp}H_{p}^{*}V^{*}\left(\beta_{vp}I_{v}-m_iI_{p}^{*}\right)
\end{align*}

In order to show LAS of the endemic equilibrium, we need to show that $\Delta(\gamma)>0$. However, due to the complexity of the formula, this will be investigated numerically. In addition, it is well known that if $\Delta(\gamma)=0$ for some values of $\gamma$, it means that the polynomial (\ref{poly}) has pure imaginary conjugate roots and one real root. However, it is straightforward to show that this real root is simply $-a_2$ (because $a_{1}a_{2}-a_{0}=0$).

In fact, if the polynomial defined in (\ref{poly}) has a pair of complex conjugate roots, $a\pm bi$ where $a,b \in \mathbb{R}$, which cross the real axis as $\gamma$ passes through $\gamma^*$, then $\Delta (\gamma)$ changes sign as $\gamma$ passes through $\gamma^*$. This can be showed very easily because we have the following relationship:
\begin{equation}
\Delta(\gamma)=-2a\left(b^{2}+r^{2}\right)-2a^{3}.
\label{Delta}
\end{equation}
Similarly, if we assume that $\Delta(\gamma)$ changes sign, then according to the previous formula $a(\gamma)$ will also changes sign too. 
However, in order to derive a Hopf Bifurcation (see, for instance, Theorem 3.4.2 in \cite{Guckenheimer-Holmes}), we have an additional property to verify, the so-called "transversality condition" that indicates that the eigenvalues cross the x-axis with a non-zero velocity. In other words
\begin{equation}
\dfrac{da}{d\gamma} (\gamma^*)=a'(\gamma^*) \neq 0
\label{transversality}
\end{equation}
However, using (\ref{Delta}), we have that 
$$
\Delta'(\gamma_i)=-2a'(\gamma^*) (b^2+r^2)
$$
such that verifying \eqref{transversality} is equivalent to verifying $\Delta'(\gamma^*)\neq 0$. 
Finally, we deduce that a Hopf Bifurcation related to the roguing parameter $\gamma$ may occur at $(EE,\gamma^*)$ if
\begin{itemize}
\item there exists a value $\gamma^*$ such that $\Delta(\gamma^*)=0$, such the Jacobian $J_{EE}$ has a simple pair of pure imaginary eigenvalues and one real negative eigenvalue, $-a_2$,
\item and $\Delta'(\gamma^*)\neq 0$.
\end{itemize}
In addition, the stability of the periodic solutions is given by the sign of the first Lyapunov coefficient of the dynamics, $l_1(EE,\gamma^*)$. If $l_1(EE,\gamma^*)< 0$ then these solutions are stable limit cycles and the Hopf bifurcation is supercritical, while if $l_1(EE,\gamma^*) > 0$ the Hopf bifurcation is subcritical \cite{kuznetsov}. $l_1$ being very difficult to obtain theoretically, we may use MatCont software \cite{Matcont} to estimate it.

We consider the parameters values provided in Table \ref{Tablevalue}, page \pageref{Tablevalue}, to illustrate Hopf Bifurcations. These values were chosen thanks to the (range of) values given in Table \ref{tabsensi}, page \pageref{tabsensi}.
\begin{table}[ht]
\centering 
\begin{tabular}{|c|c|c|c|c|c|c|c|c|c|c|c|}
\hline
$r$ & $\phi$ & $m_h$ & $m_i$ & $\beta_{vp}$ & $\gamma$ & $\alpha_v$ & $\mu_1$ & $\mu_2$ & $\beta_{pv}$ & $k_p$ & {$\delta$}\\
 \hline
 $0.04$ & $0$ & ${0.01}$ & ${0.01}$ & $0.008$ & $[0,1]$ & $0.2$ & $0.12$ & $0.0002$ & ${0.02}$ & $0.005$ & $0$\\
 \hline 
\end{tabular}
\caption{Parameters values}
\label{Tablevalue}
\end{table}

In Table \ref{tableHB}, page \pageref{tableHB}, we present Hopf bifurcation results related to different values of $k_p$. In fact, it is clear that if $k_p\geq k^*_p$, then there is no Hopf Bifurcation. 
\begin{table}[ht]
\begin{center}
\begin{tabular}{|c|c|c|c|c|c|c|}
 \hline
 $k_p$ & & $H_i$ & $I_p$ & $I_v$ & $\gamma_i^*$ & $l_1$ \\
 \hline
$0.0025$& $H_2$ & 0.561818 & 0.097525 & 3.863322 & 0.208952 & 
$-2.953532e^{-05}$ \\
 \hline 
 & $H_1$ & 0.030780 & 0.113949 & 4.506589 & 0.045791 &
 $-1.590897e^{-05}$ \\
 \hline
 $\bf{0.005}$& $H_2$ & 0.524541 & 0.090960 & 3.605623 & 0.195185 & $-3.069432e^{-05}$ \\
 \hline 
 & $H_1$ & 0.031325 & 0.106193 & 4.203102 & 0.043544 &
 $-1.620631e^{-05}$ \\
 \hline
 ${0.0072}$& $H_2$ & 0.490470 & 0.085230 & 3.380376 & 0.182667 & $-1.659246e^{-05}$ \\
 \hline 
 & $H_1$ & 0.031934 & 0.099374 & 3.935864 & 0.041605 &
 $-1.659246e^{-03}$ \\
 \hline
\end{tabular}
\end{center}
\caption{Hopf Bifurcation Points coordinates and First Lyapunov coefficient estimate for several harvesting rates (based on Table \ref{tabsensi}, page \pageref{tabsensi}).}
\label{tableHB}
\end{table}

\begin{figure}[ht]
\begin{center}
 \includegraphics[width = 1.0\textwidth]{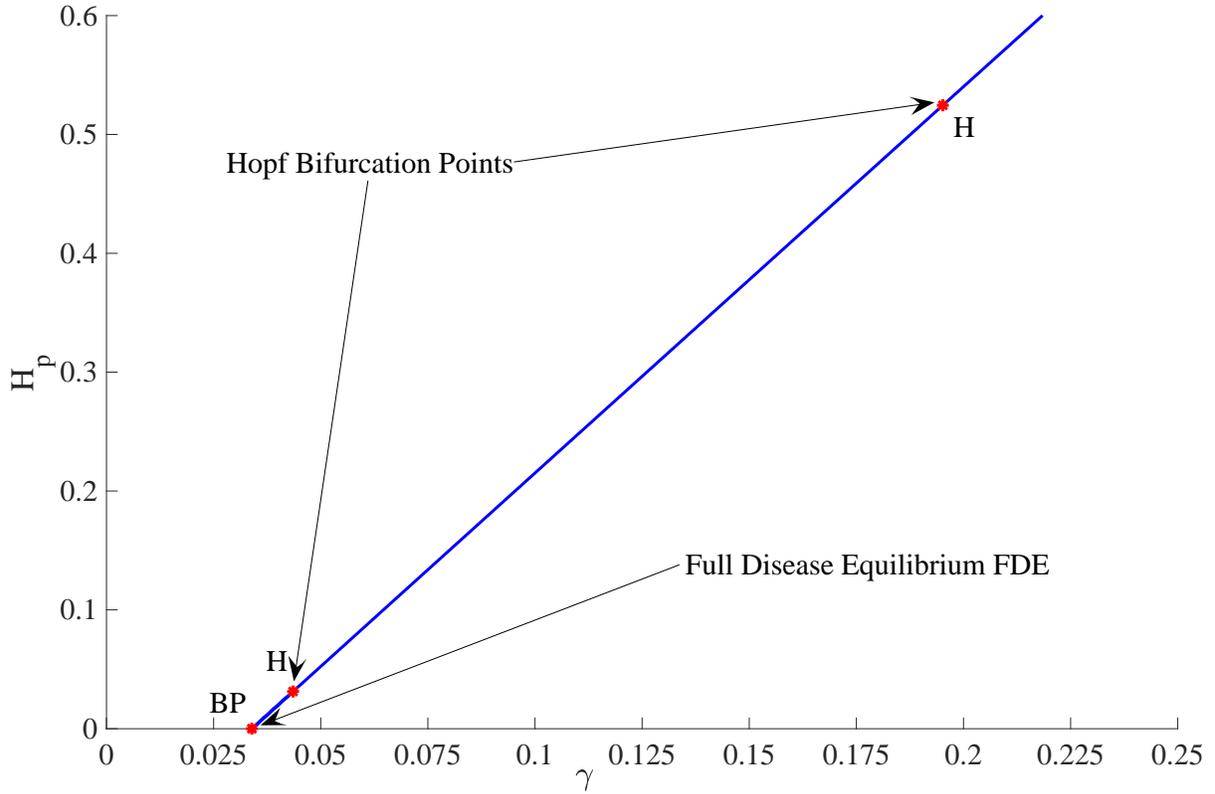}
  \caption{Bifurcation analysis made with MatCont - $k_p=0.005$}
\label{figMatcont}
\end{center}
\end{figure}

To illustrate the Hopf Bifurcation property, we focus on  $k_p=0.005$, in Table \ref{tableHB}, when $\delta=0$, for simplicity. In Fig. \ref{figMatcont}, page \pageref{figMatcont}, we show the bifurcation analysis of system \eqref{model2} done with MatCont \cite{Matcont}, when $\gamma \in [0.02,0.12]$: we show that two values, $\gamma_1^*$ and $\gamma_2^*$, exist where a Hopf bifurcation occurs. 

\begin{figure}[ht]
\begin{tabular}{c@{\hspace{0.5cm}}c}
 \includegraphics[width = 0.5\textwidth]{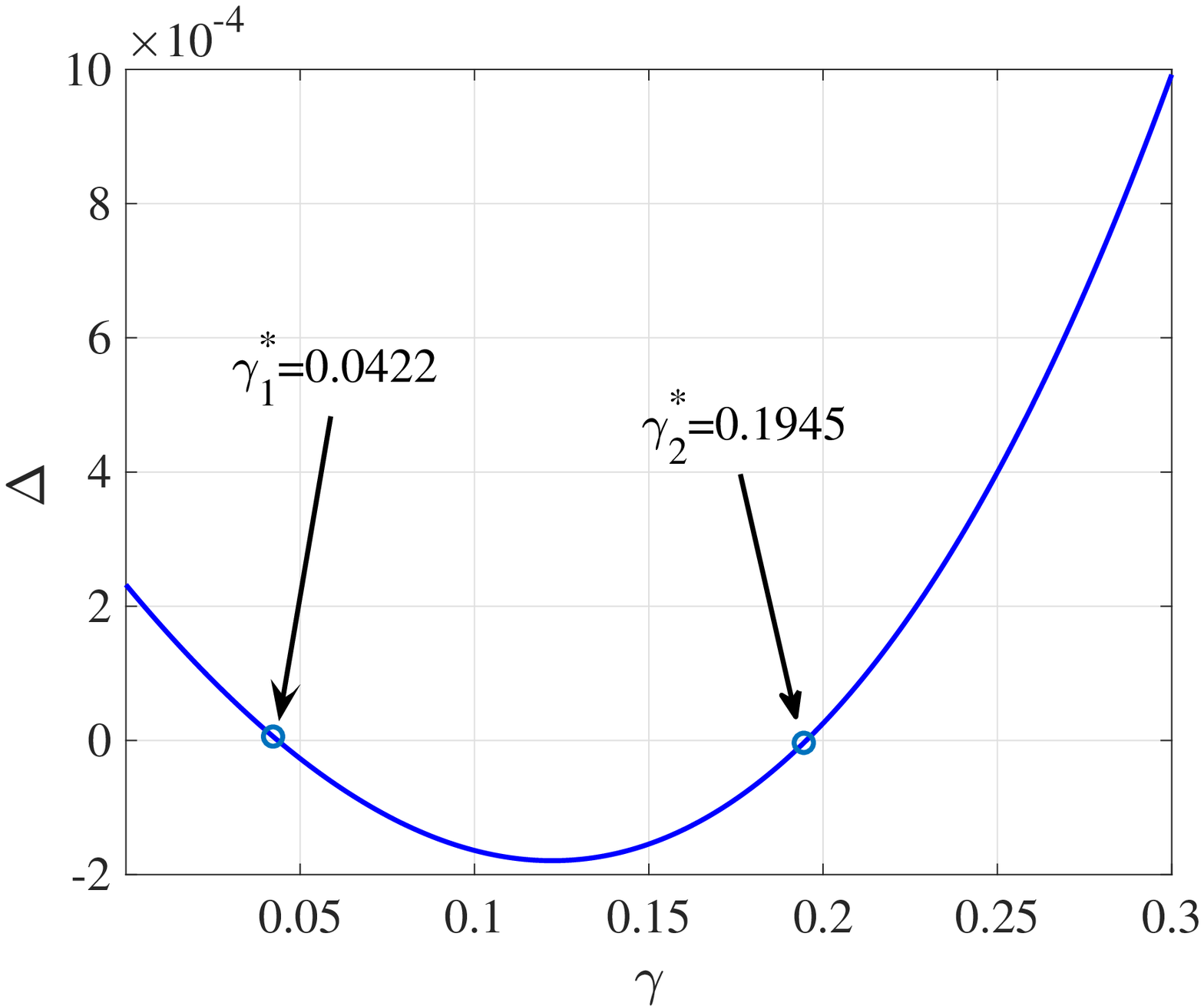}& \includegraphics[width = 0.5\textwidth]{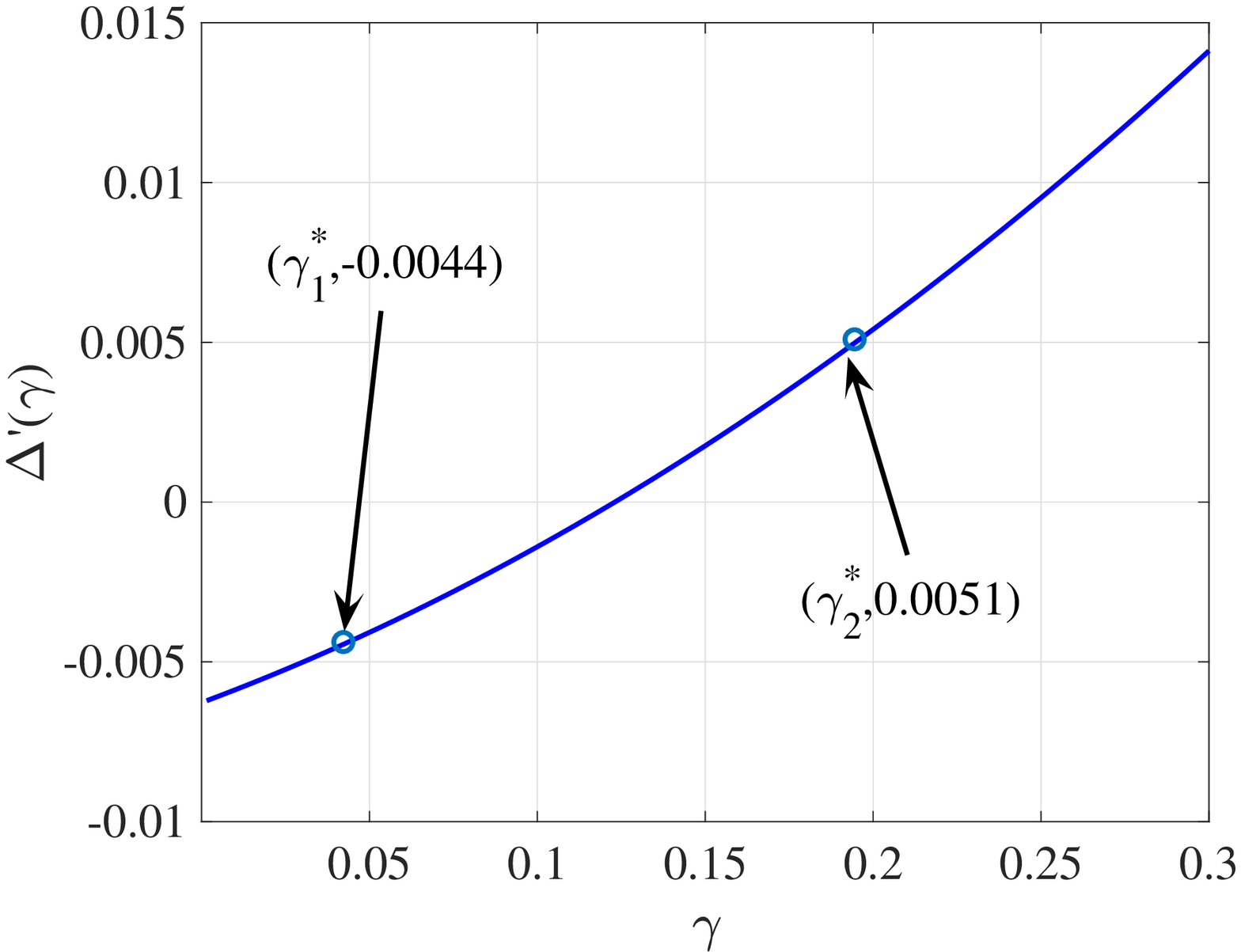}\\
(a) \; $\Delta$ with $\gamma$. &  (b) \; $\Delta'$ with $\gamma$. \\
  \end{tabular}
  \caption{Illustration that the second Rough-Hurwitz condition can give imaginary eigenvalues ($k_p=0.005$).}
\label{fig14}
\end{figure}

However, these two values can be obtained by solving $\Delta(\gamma^*)=0$. This is illustrated in Figures \ref{fig14}, page \pageref{fig14}, and \ref{fig11}, page \pageref{fig11}.

\begin{figure}[ht]
\begin{center}
 \includegraphics[width = 0.8\textwidth]{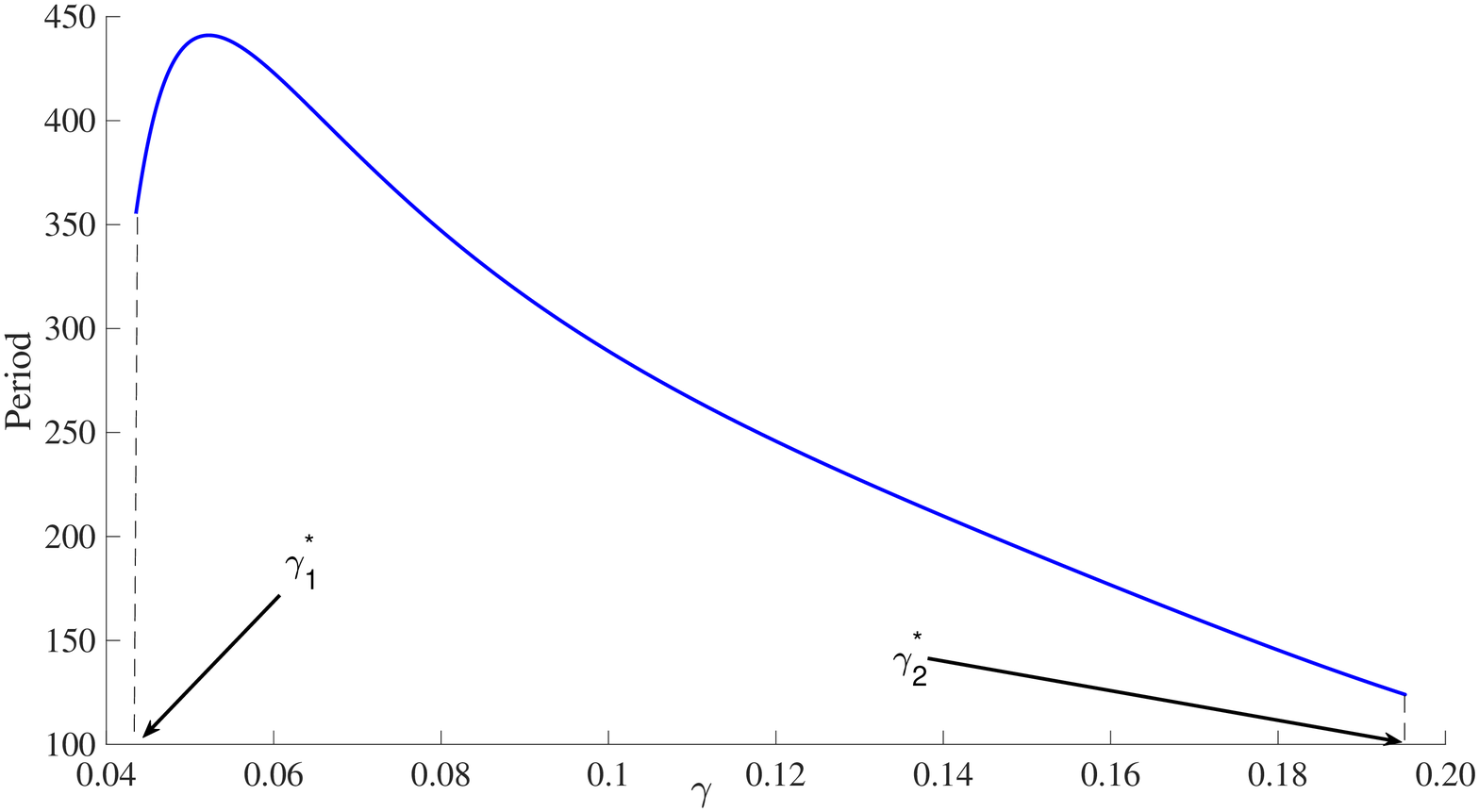}
  \caption{Period estimate of the periodic solution according to $\gamma \in [\gamma_1^*,\gamma_2^*]$, using MatCont, for $k_p=0.005$.}
\label{period}
\end{center}
\end{figure}

Last but not least, since we know that the Hopf bifurcation points are \tcr{supercritical}, we can estimate the periods for different values of $\gamma \in [\gamma_1^*,\gamma_2^*]$, using MatCont. See Fig. \ref{period}, page \pageref{period}. Surprisingly, the period belongs to a large interval, i.e. in $[130, 445]$. This period might certainly be linked to the crop growth parameters. However, since the period is large, except for values of $\gamma$ closed from $\gamma_2^*$. However, the maximum period, $445$ days, is reached at $\gamma=0.05$. This period is almost $15$ months, barely impossible to detect it in the field.

\section{Numerical simulations}\label{sec:numerical}
\label{section5}
\tcm{For the sensitivity analysis and the numerical simulations, we first discuss the parameters values given in Table \ref{tabsensi}, page \pageref{tabsensi}} . Some of them are based on the values used in \cite{Keating1982,holt,Jeger2004}, and some of them are estimated based on the available knowledge on the virus and the cassava crops. In fact, despite so much work and published papers on cassava and CMV, parameter estimation has always been a critical issue. Thus, after carefully checking the literature on cassava and CMD, we have tried to provide "realistic" estimates for all parameters used in our model. It is important to note that here we consider cassava biomass and not individual plants, hence some adjustments to the literature values were necessary.

In fact, it is not easy to get information about cassava growth: among all the work done on cassava, surprisingly, only one study from Australia was found about cassava biomass growth, \cite{Keating1982}. These are not exactly the same environmental conditions found in Africa, but in any case, at the plants level, this study shows that a logistic equation can represent well the (dry) biomass growth. We will consider the values obtained in \cite{Keating1982} to estimate the range of values of $r$ and $m_{h}$. \tcm{In Africa, the range of planting density can be rather large. However, for the sake of simplicity, we will consider a standard density of $10000$ plants per ha, or $1$ plant per $m^2$.}


According to (FAOSTAT2019), the average productivity between 2010 and 2017 is around $9$ tons per ha in Africa, this leads to a mean productivity of $0.9$ kg per $m^{2}$. Then assuming an harvest index (root weight divided by total plant weight) between $0.5$ and $0.7$, these leads to a total biomass weight per square meter between $1.3$kg and $1.8$ kg. Typically, harvest can start $8$ months (250 days) after planting, so that the harvest rate $k_p$ ranges from $1.3/250$ to $1.8/250$, that is $k_{p}\in[0.0052,0.0072]$ roughly. If the harvest
starts later, say $500$ days after planting, then $k_{p}\in[0.0026,0.0036]$, so that we can choose the interval $[0.0025,0.0072]$ for the harvesting rate in our model.

Last, but not least, the estimate of acquisition and inoculation rates is also another issue. Only two old studies have been done (see for instance \cite{Fargette1990}) on the transmission efficiency and it is in general very low (from $0.17\%$ to $2\%$) for the vectors. Surprisingly, we didn't find information related to \tcm{the number of plants visited by vectors, or the number of feeding bites, such that estimate of acquisition and inoculation rates is rather difficult. According to the literature, it seems that the transmission rate, from the infected vector to the susceptible host, $\beta_{vp}$, is rather low; in contrary, thanks to the fact that the feeding time is long (couple of hours), the inoculation rate from the infected plant to the susceptible vector, $\beta_{pv}$, can be high. Thus, without detailed information available in the literature, we will use the estimates of $k_1$ and $k_2$ given in \cite{holt}, to estimate $\beta_{vp}$ and $\beta_{pv}$. However, due to the fact that they have been estimated for individual plants and not for biomass we have to divide by the total biomass weight per square meter given above, since we consider a density of one individual per square meter. Thus $\beta_{vp} \in [0.0011, 0.0246]$, while for $\beta_{pv}$, we will not consider the same range, but a larger range, say $[0.0056,0.23]$.}

\subsection{Sensitivity analysis of model (\ref{model1})}
To gain insight into the correct strategies for control of the crop vector-borne disease as described by model \eqref{model1}, we perform a sensitivity analysis.

The most important parameter is the basic reproduction number, $\Rov$. In the current context it represents the amount of new infections per unit of plant biomass (vector) by the introduction of one unit of infected plant biomass (vector).
In this section we aim to find out how the basic reproduction ratio respond to changes in the selected parameters. Mathematically, the sensitivity of $\Rov$ with respect to a parameter $p$ is given by
\[
\mathscr{F}^{\Rov}_p=\dfrac{\partial \Rov}{\partial p}.
\]
However, we will consider the sensitivity index. This measures the change in $\Rov$ with respect to the percentage change in parameter $p$.
The normalised sensitivity index of a variable $\Rov$, that depends differentiably on a parameter $p$, is defined as
\[
\mathscr{E}^{\Rov}_p=\dfrac{\partial \Rov}{\partial p}\frac{p}{\Rov}.
\]
The sensitivity index of $\Rov$ with respect to the parameter $p$ is positive if $\Rov$ is increasing with respect to
$p$ and negative if $\Rov$ is decreasing with respect to $p$. For convenience, and following the discussion in Remark \ref{remark1}, we present the local sensitivity analysis on $\Rov$.
Straightforward calculation leads to the following results:

\begin{align*}
& \mathscr{E}^{\Rov}_r=\frac{\gamma r}{[(m_i-m_h)P^*+\gamma]P^*m_h(1+\phi V^*)},\quad 
 \mathscr{E}^{\Rov}_\phi= -\frac{\gamma\phi}{[(m_i-m_h)P^*+\gamma]P^*m_h(1+\phi V^*)^2}, \\
& \mathscr{E}^{\Rov}_{k_p}= -\frac{\gamma\phi}{[(m_i-m_h)P^*+\gamma]P^*m_h}, \quad 
 \mathscr{E}^{\Rov}_{m_h}= -\frac{\gamma\phi(r_{p }(V^*)-k_p)}{[(m_i-m_h)P^*+\gamma]P^*m_h^2}, \\
& \mathscr{E}^{\Rov}_{m_i}=-\frac{P^*m_i}{(m_i-m_h)P^*+\gamma}, \quad  \mathscr{E}^{\Rov}_{\alpha}=\frac{\mu_1}{\mu_2V^*},\quad 
 \mathscr{E}^{\Rov}_{\mu_1}=-\frac{\mu_1}{\mu_2V^*}, \\
& \mathscr{E}^{\Rov}_{\mu_2}=-1, \quad 
 \mathscr{E}^{\Rov}_{\beta_{pv}}= 1, \quad
 \mathscr{E}^{\Rov}_{\beta_{vp}}= 1, \\
& \mathscr{E}^{\Rov}_{\gamma}=-\frac{\gamma}{(m_i-m_h)P^*+\gamma}. 
\end{align*} 

\begin{figure}[ht]
\begin{tabular}{c@{\hspace{0.5cm}}c}
  \includegraphics[width = 0.5\textwidth]{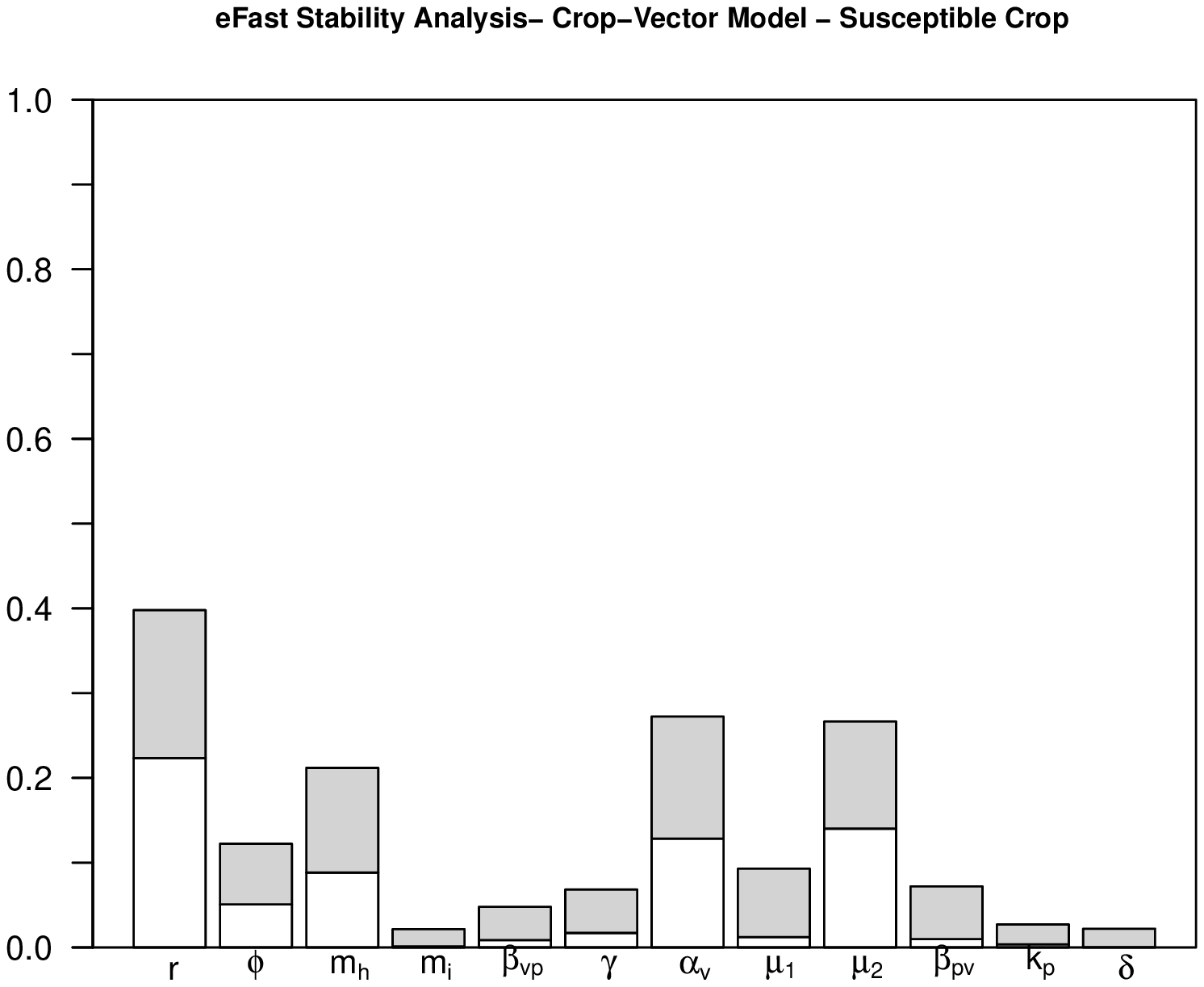} &
  \includegraphics[width = 0.5\textwidth]{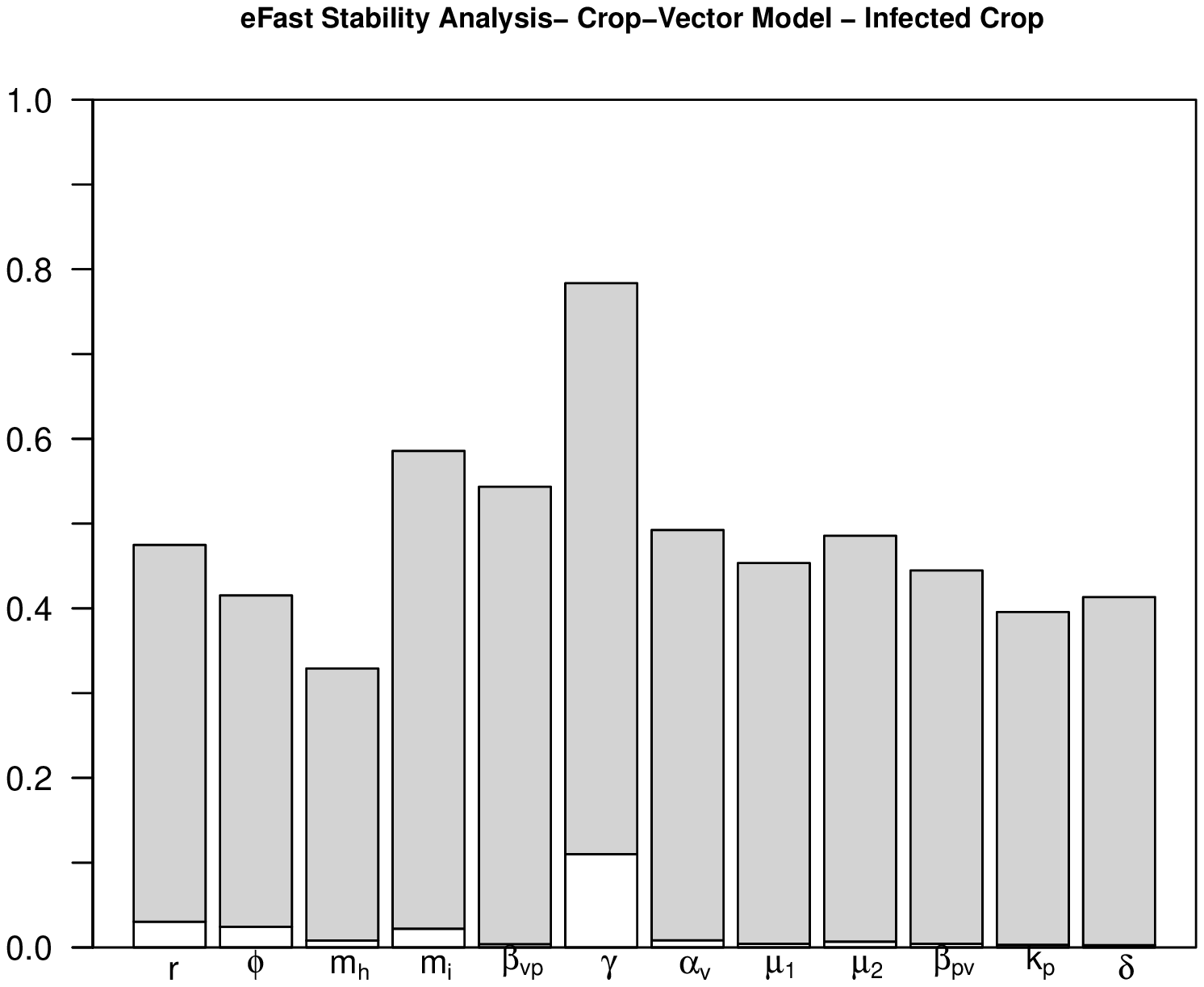} \\
 \includegraphics[width = 0.5\textwidth]{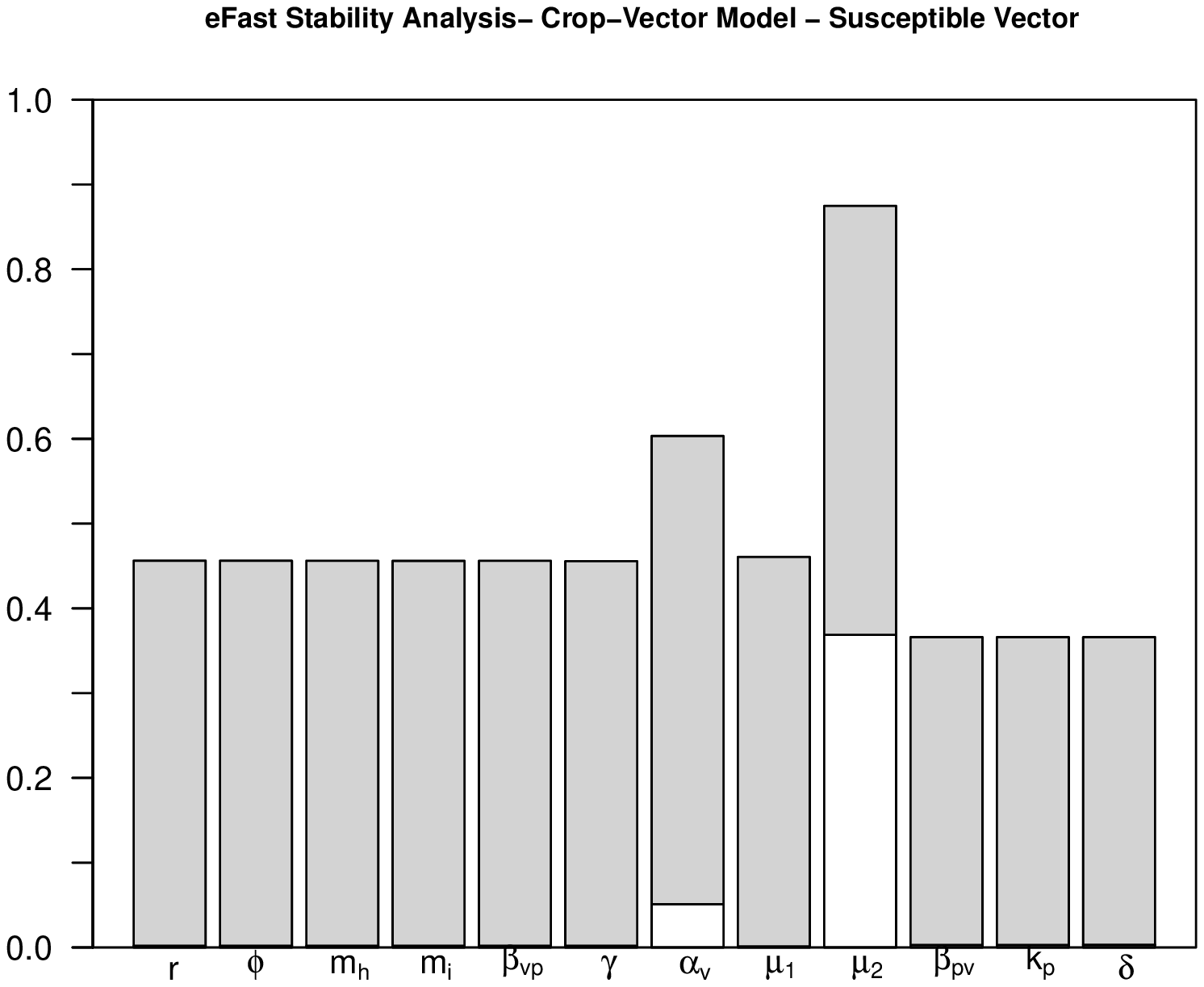} &
 \includegraphics[width = 0.5\textwidth]{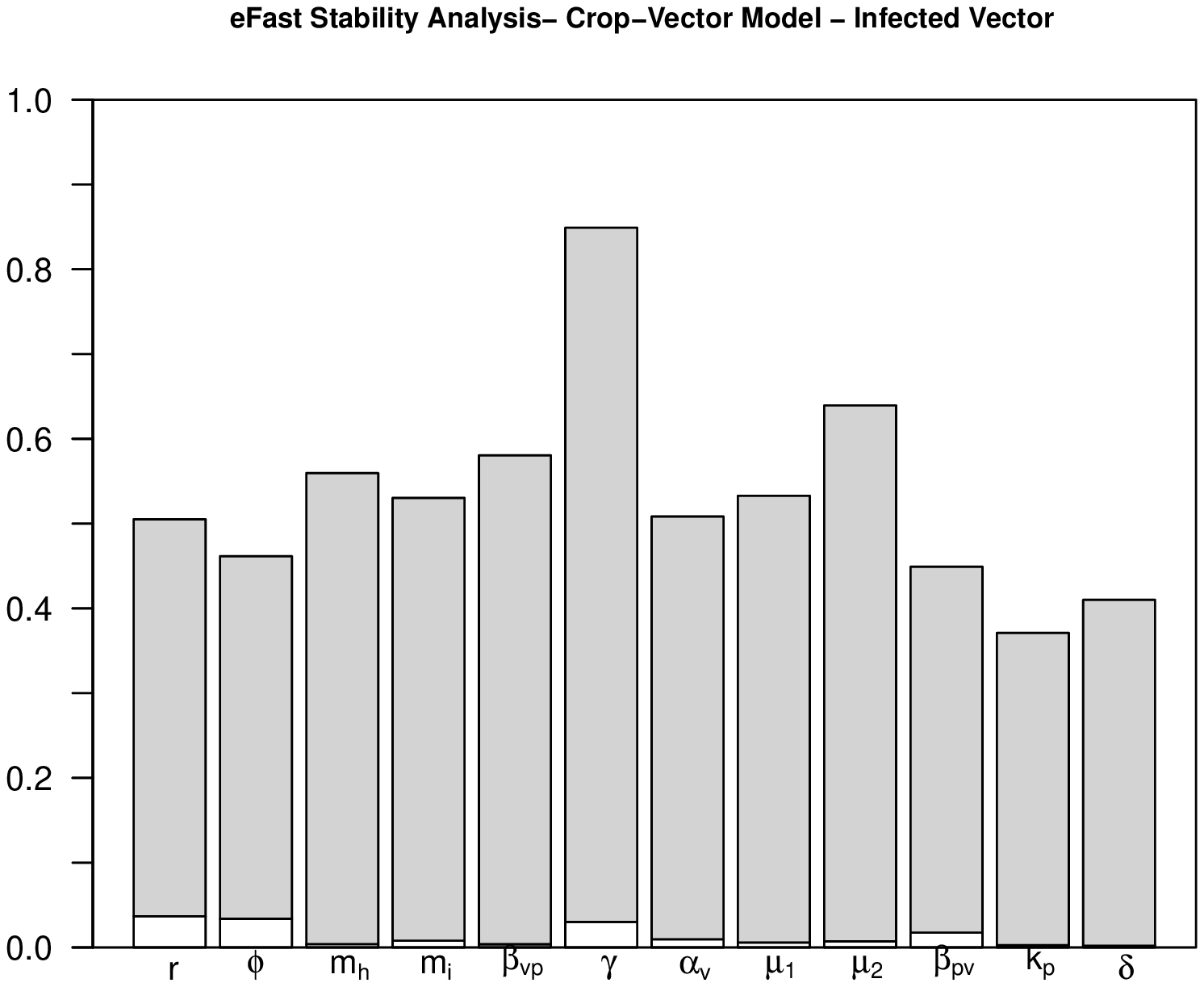}
\end{tabular}
\caption{e-FAST Sensitivity analysis. White bar:first-order effects; Sum of white and grey bars: total effect.}
\label{efast}
\end{figure}

\begin{figure}[ht]
\begin{tabular}{c@{\hspace{0.5cm}}c}
 \includegraphics[width = 0.5\textwidth]{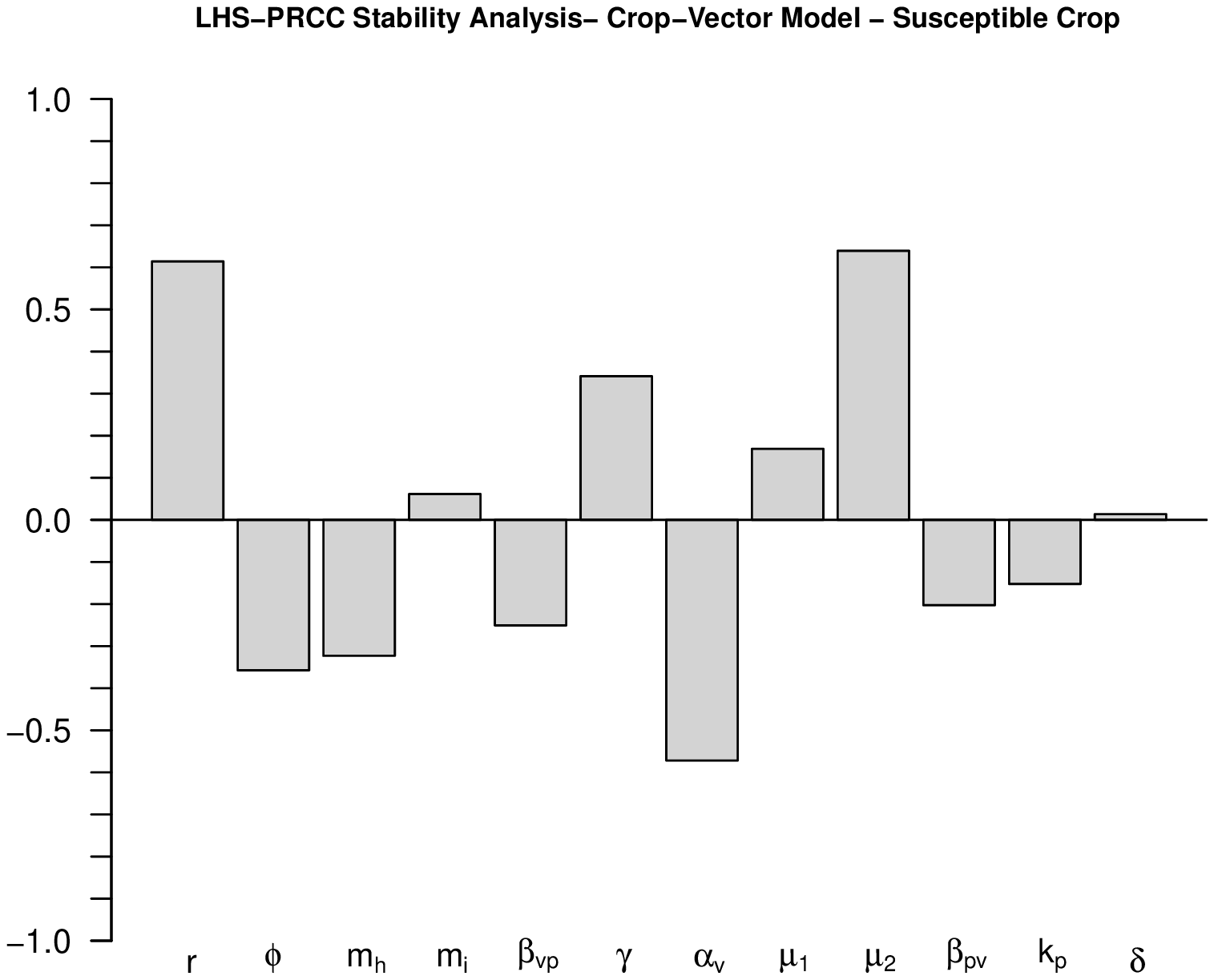} &
 \includegraphics[width = 0.5\textwidth]{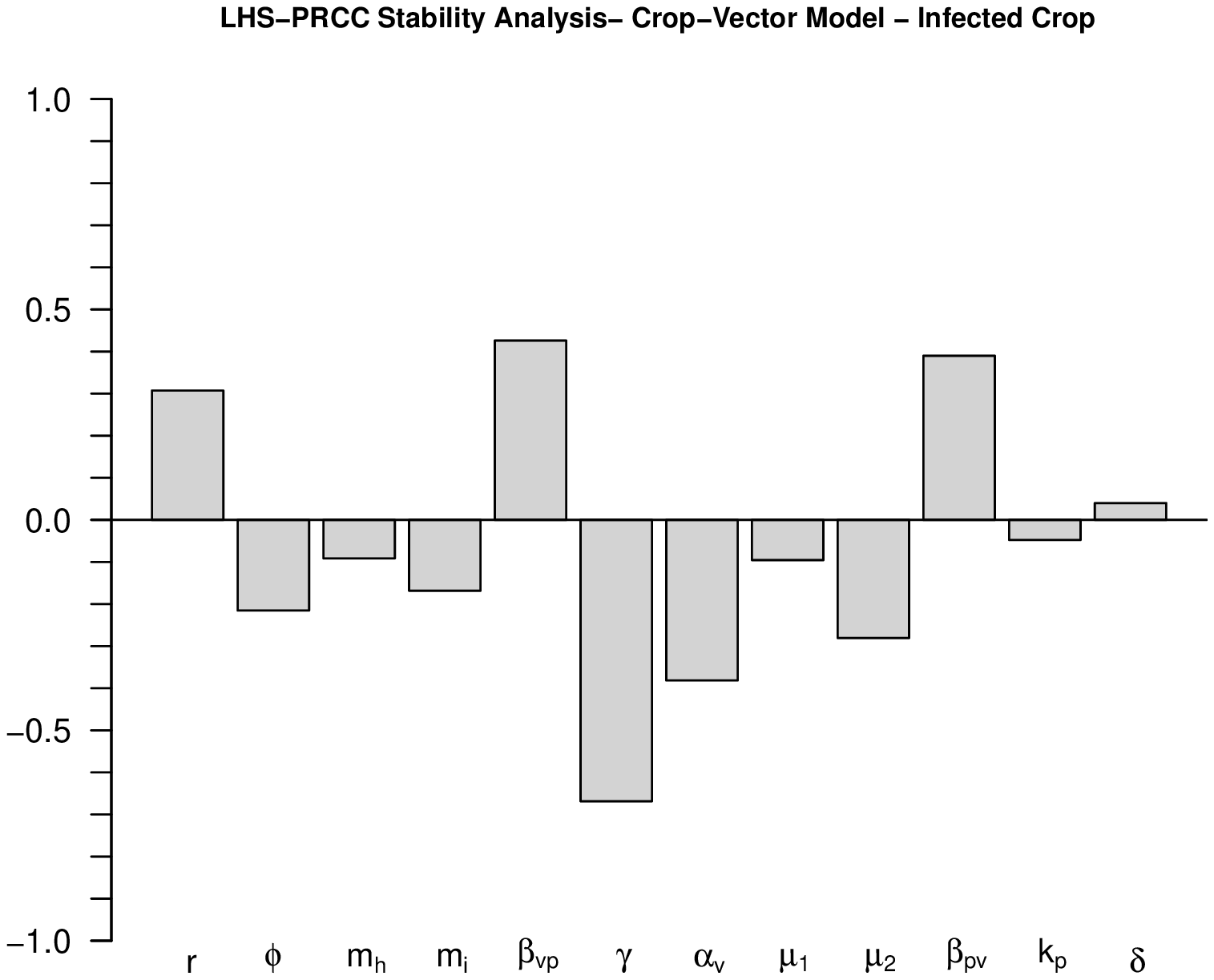} \\
 \includegraphics[width = 0.5\textwidth]{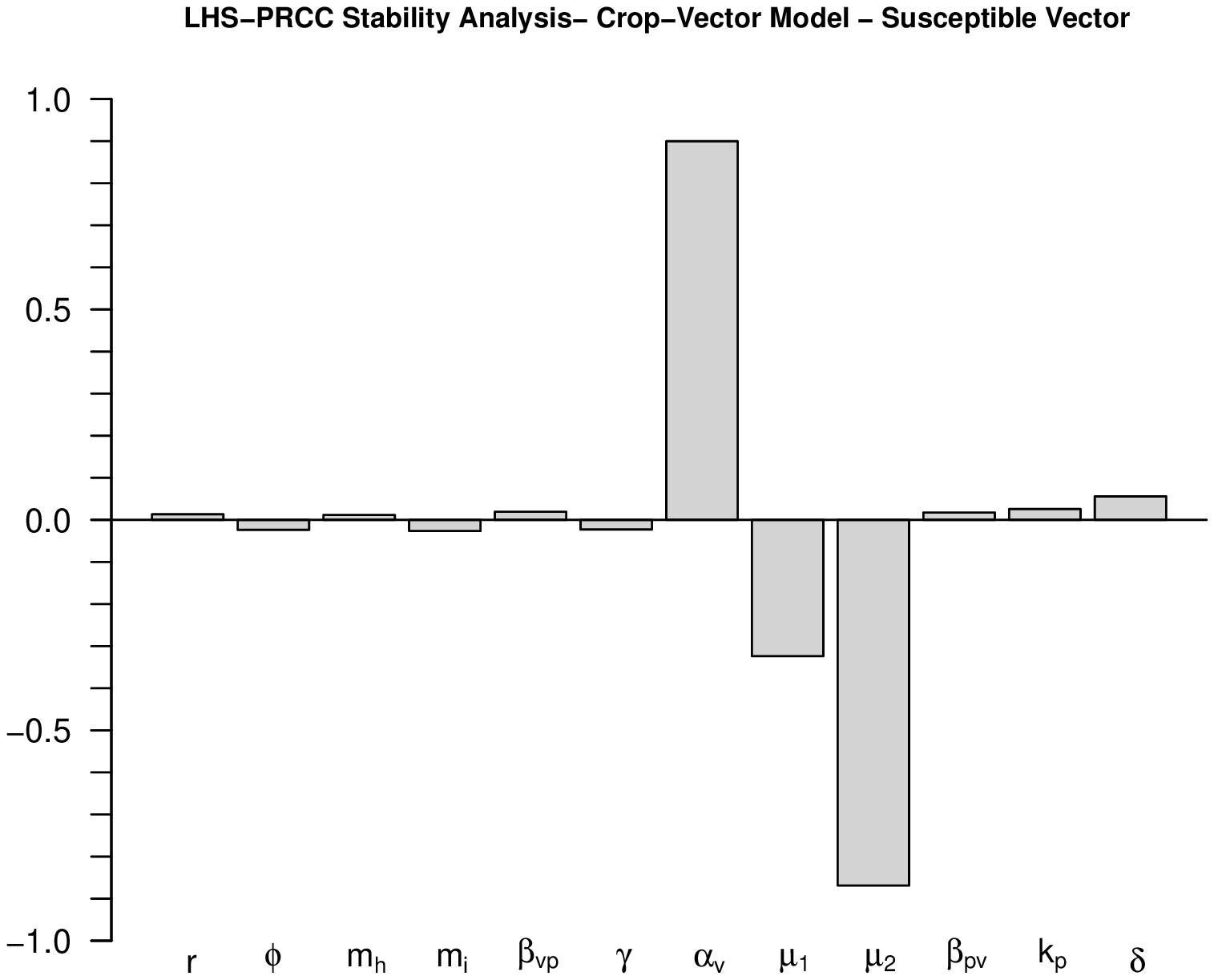} &
 \includegraphics[width = 0.5\textwidth]{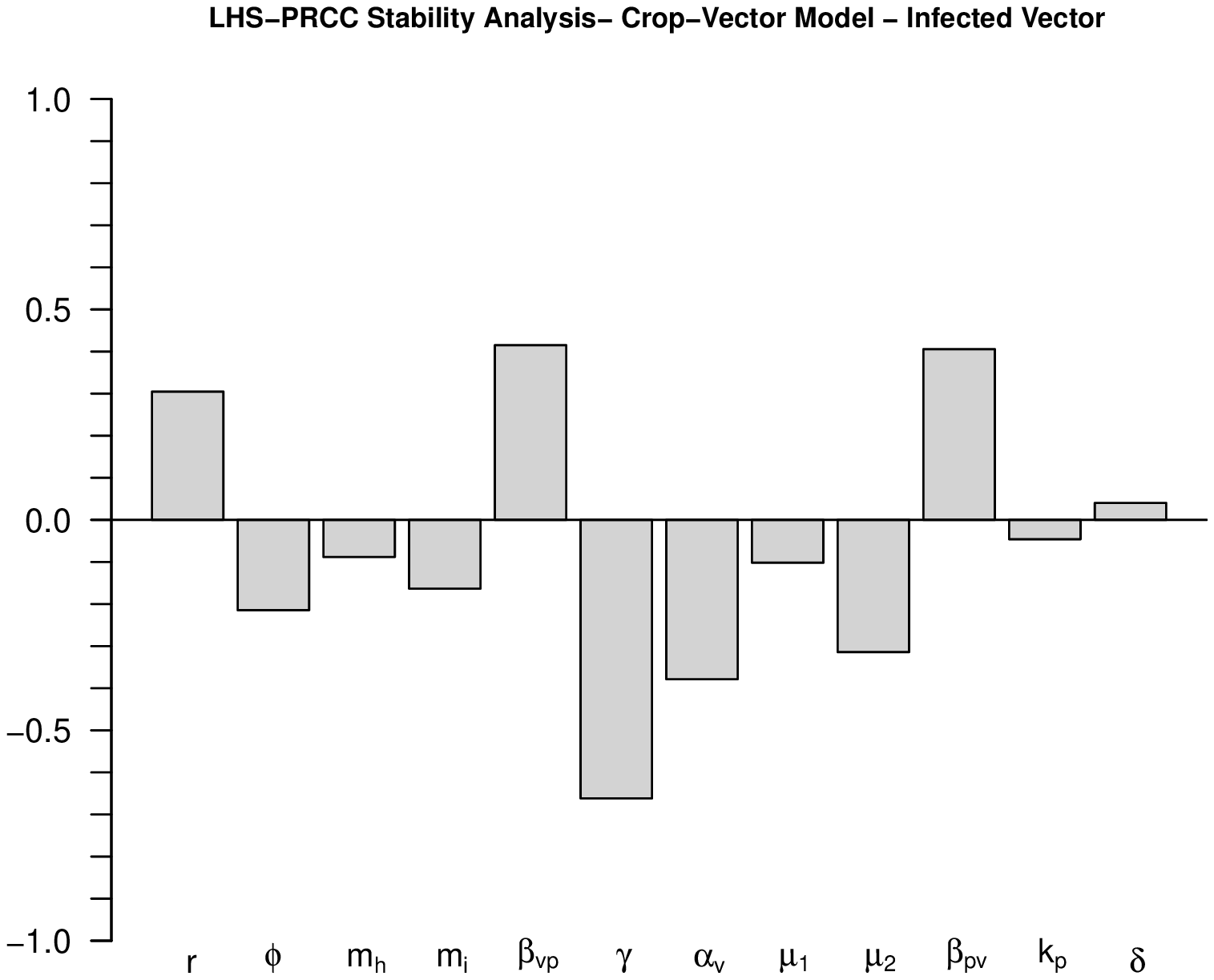}
 \end{tabular}
  \caption{LHS-PRCC Sensitivity analysis.}\label{PRCC}
\end{figure}

However, the sensitivity index gives us only partial information because we consider only the sensitivity for one parameter only and also only on $\Rov$. We will now consider an additional sensitivity analysis considering that all parameters are changing. We will also focus on the system's variables $H_p$, $I_p$, $S_v$, and $I_v$. Using Table \ref{tabsensi}, page \pageref{tabsensi}, we derive some global sensitivity analysis using two well-known methods: the eFast and the LHS-PRCC methods. The eFast method given in Fig.\ref{efast}, page \pageref{efast}, highlights first-order effects (main effects) and total effects (main and all interaction effects) of the parameters on the Model Outputs. We also derive a LHS-PRCC sensitivity analysis given in Fig.\ref{PRCC}, page \pageref{PRCC}. LHS-PRCC stands for Latin Hypercube Sampling and PRCC for Partial rank correlation coefficient. These two methods give complementary information. Indeed the PRCC method provides mainly information about how the outputs are impacted if we increase (or decrease) the inputs of a specific parameter while the eFast indicates which parameter uncertainty has the greatest impact on the output variability (see for instance \cite{marino2008} for further explanations). Clearly here, the LHS-PRCC method provides the most interesting sensitivity analysis in terms of the contribution of each parameter, that may depend on the chosen variable. However, some parameters, related to the whiteflies dynamics ($\alpha_v$, $\mu_2$) or the transmission ($\beta_{vp}$ and $\beta_{pv}$ or the plant growth ($r$, $\phi$, $\gamma$) may have a strong effect on the system dynamics. \tcm{The values for the intervals used for the sensitivity analysis are given in Table \ref{tabsensi}, page \pageref{tabsensi}.}

\begin{figure}[ht]
\centering 
 \includegraphics[scale=0.7]{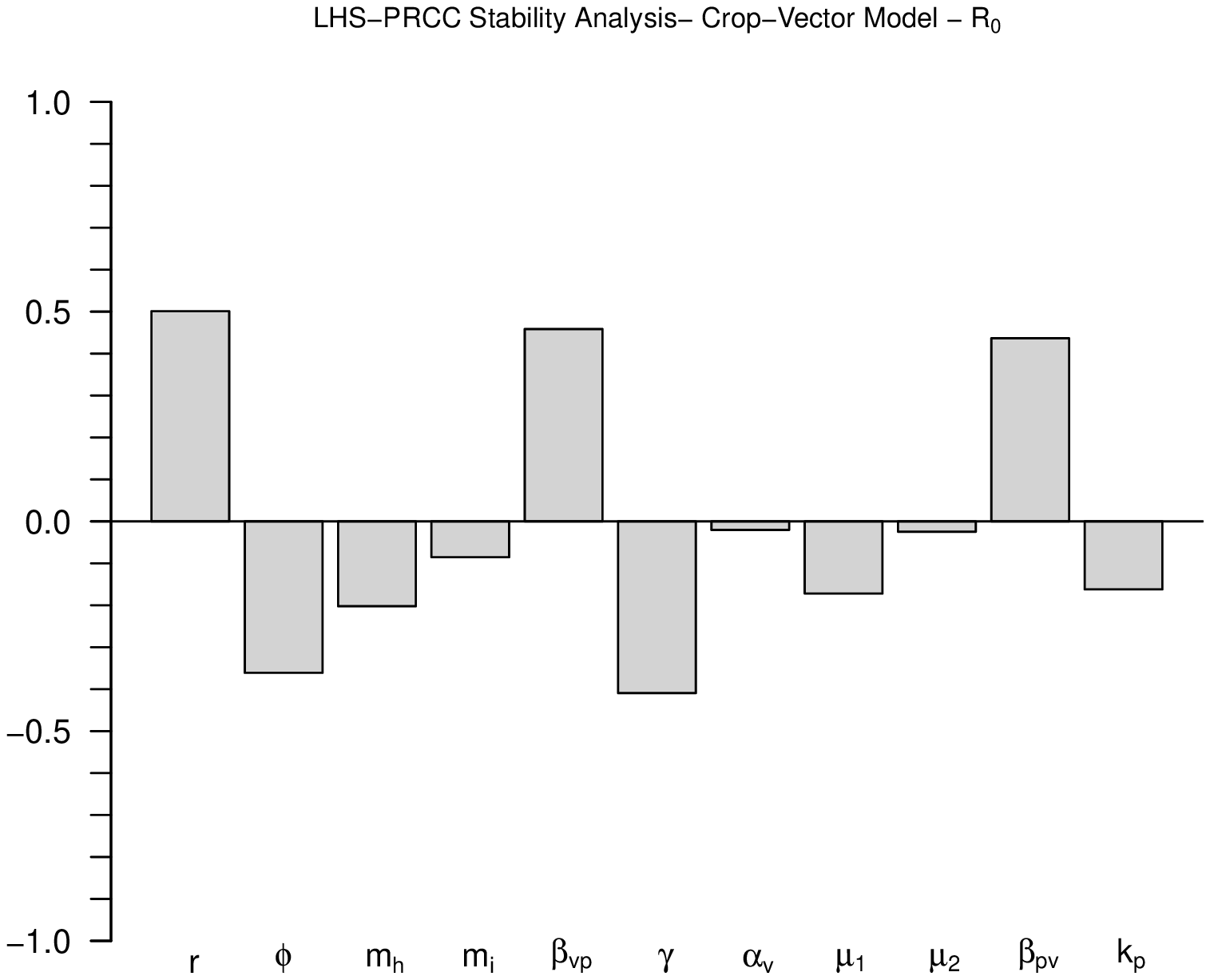}
   \caption{LHS-PRCC Sensitivity analysis of the Basic reproduction numbers $\Ro$.}\label{PRCC-R0}
\end{figure}

\begin{figure}[ht]
\centering
  \includegraphics[scale=0.8]{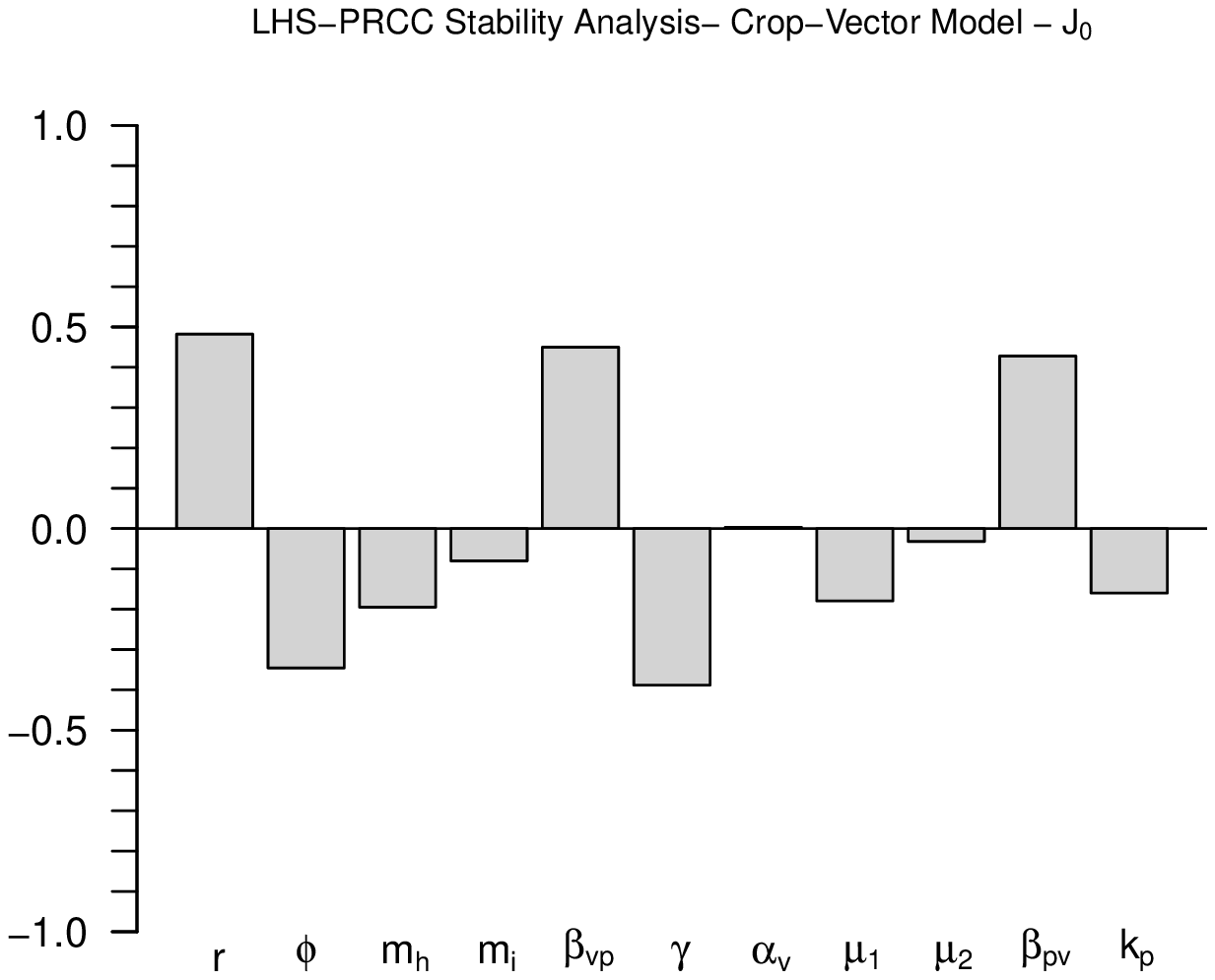}
  \caption{LHS-PRCC Sensitivity analysis of the Basic reproduction number $\mathcal{J}_{0}$.}\label{PRCC-R0v}
\end{figure}

In Figs. \ref{PRCC-R0}, page \pageref{PRCC-R0}, and \ref{PRCC-R0v}, page \pageref{PRCC-R0v}, we show the LHS-PRCC sensitivity analysis of both Basic reproduction numbers, $\Ro$ and $\mathcal{J}_{0}$. Since they are mathematically equivalent, it make sense that their sensitivity analysis are almost similar. The results are in agreement with what we expect: when some parameters increase, then either $\Ro$ or $\mathcal{J}_{0}$ increases or decreases. The thresholds are more sensitive to some parameters, like $\beta_{vp}$, $\beta_{pv}$, $r_p$ or $\phi$. However, we can see that the roguing parameter, $\gamma$, has a strong negative effect, like $k_p$, the harvest rate. As we have seen earlier, very little is know about the values taken by $\beta_{vp}$ and $\beta_{pv}$, such that additional studies would be welcome to conduct.


\subsection{Numerical simulations}

The above system of ordinary differential equations is highly nonlinear, hence in this section we present numerical simulations to support the results of the previous sections. In addition, we also present simulations to illustrate the model behavior with respect to the given parameters. We begin with the basic model where all parameters are chosen according to the base line values provided in Table \ref{tabsensi}, page \pageref{tabsensi}. The system is integrated using MatLab's \emph{ode solvers} and we adjust the solver's \emph{RelTol} and \emph{AbsTol} until the simulations have converged. The rest of the simulation were then performed using these baseline tolerance values.

The following simulations are done using parameters values given in Table \ref{Tablevalueb}, page \pageref{Tablevalueb}, and chosen using Table \ref{tabsensi}, page \pageref{tabsensi}.



\begin{table}[ht]
\begin{tabular}{|c|c|c|c|c|c|c|c|c|c|c|c|}
\hline
$r$ & $\phi$ & ${m_h}$ & ${m_i}$ & $\beta_{vp}$ & $\gamma$ & $\alpha_v$ & $\delta$& $\mu_1$ & $\mu_2$ & $\beta_{pv}$ & $k_p$ \\
 \hline
 $0.04$ & $0$ & ${0.01}$ & ${0.01}$ & $0.008$ & $[0,1]$ & $0.2$ & $[0,1]$ & $0.12$ & $0.0002$ & ${0.02}$ & $0.005$\\
 \hline 
\end{tabular}
\caption{Parameters values with $\delta\geq 0$.}
\label{Tablevalueb}
\end{table}

\begin{figure}[ht]
\begin{tabular}{c@{\hspace{0.5cm}}c}
 \includegraphics[width = 0.5\textwidth]{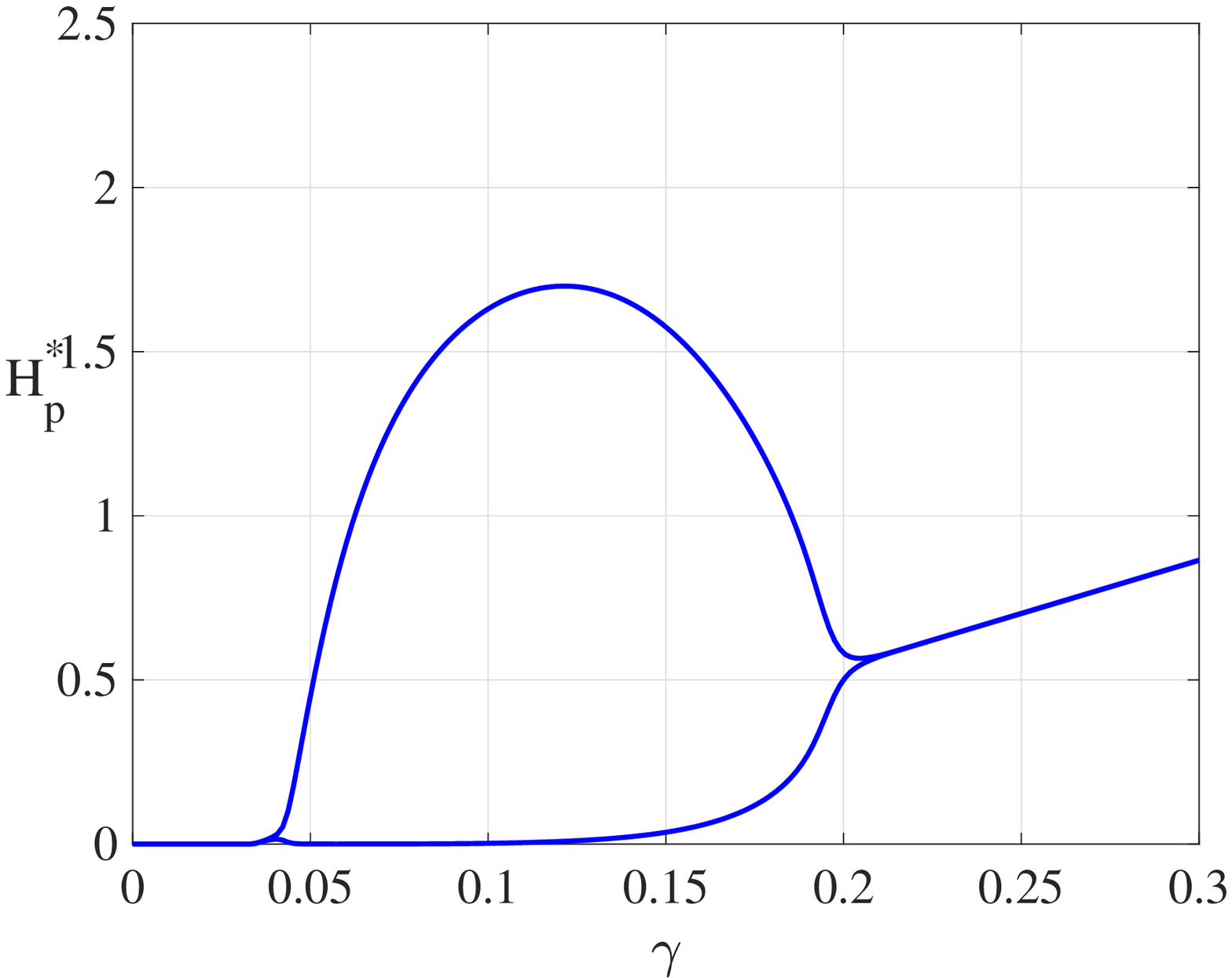}& \includegraphics[width = 0.5\textwidth]{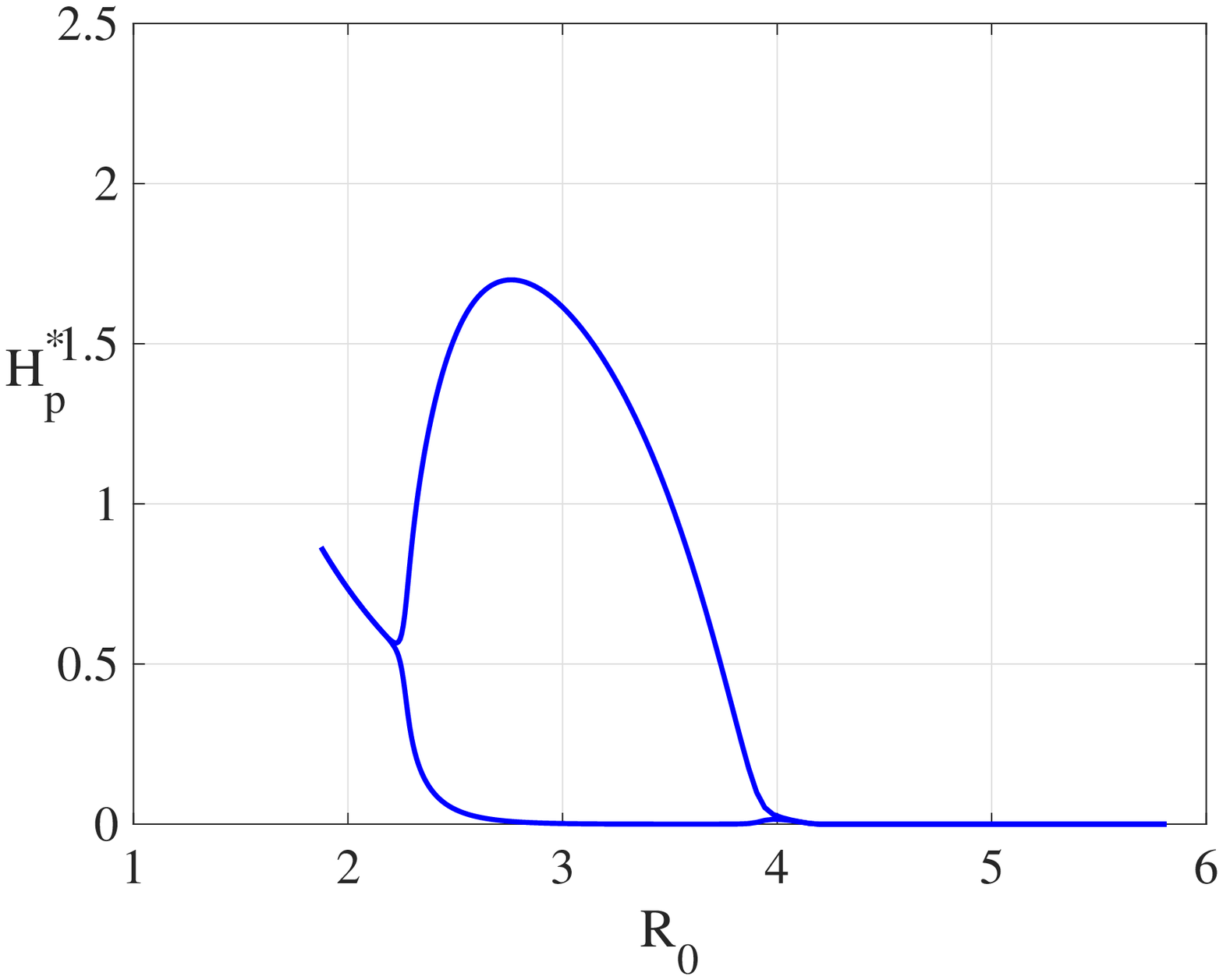}\\
(a) \; Bifurcation with respect to $\gamma$. &  (b) \; Bifurcation with respect to $\Ro$. \\
  \end{tabular}
  \caption{Bifurcation diagram showing the maxima and minima of Healthy plant biomass - $k_p=0.005$.}
\label{fig11}
\end{figure}

In particular, when $\delta=0$, Fig. \ref{fig11}, page \pageref{fig11}, explores the impact of the roguing parameter, $\gamma$ and also of $\R_0$ on the Healthy crop biomass, through the Hopf bifurcation. Thus, we consider either $\gamma$ (Fig. \ref{fig11}(a)) or $\R_0$ (Fig. \ref{fig11}(b)) as bifurcation parameter and the vertical axis shows the values of the Healthy plant biomass. In Fig. \ref{fig11} (a), an increase in $\gamma$, correspond to a large amount of healthy plant biomass at equilibrium. Then, with $\R_0$ we obtain the opposite: when $\R_0$ is large, then $H_p=0$ and the Full Disease equilibrium is reached. 

Clearly, thanks to the plot of the stability function $\Delta$ in Fig. \ref{fig14}, the endemic equilibrium loses stability as $\R_0$ decreases when oscillatory solutions appear.

\begin{figure}[ht]
\begin{tabular}{c@{\hspace{0.5cm}}c}
 \includegraphics[width = 0.5\textwidth]{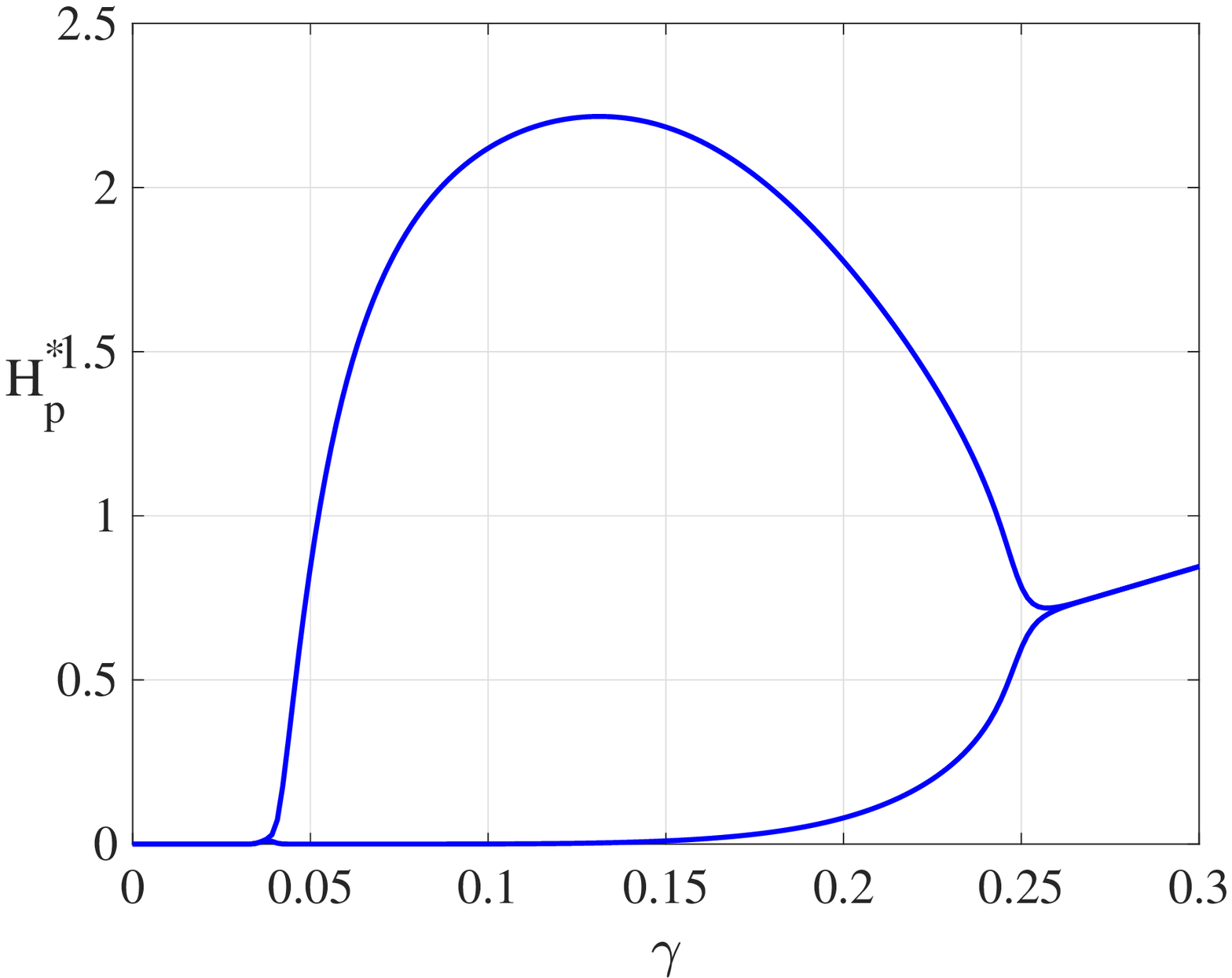}& \includegraphics[width = 0.5\textwidth]{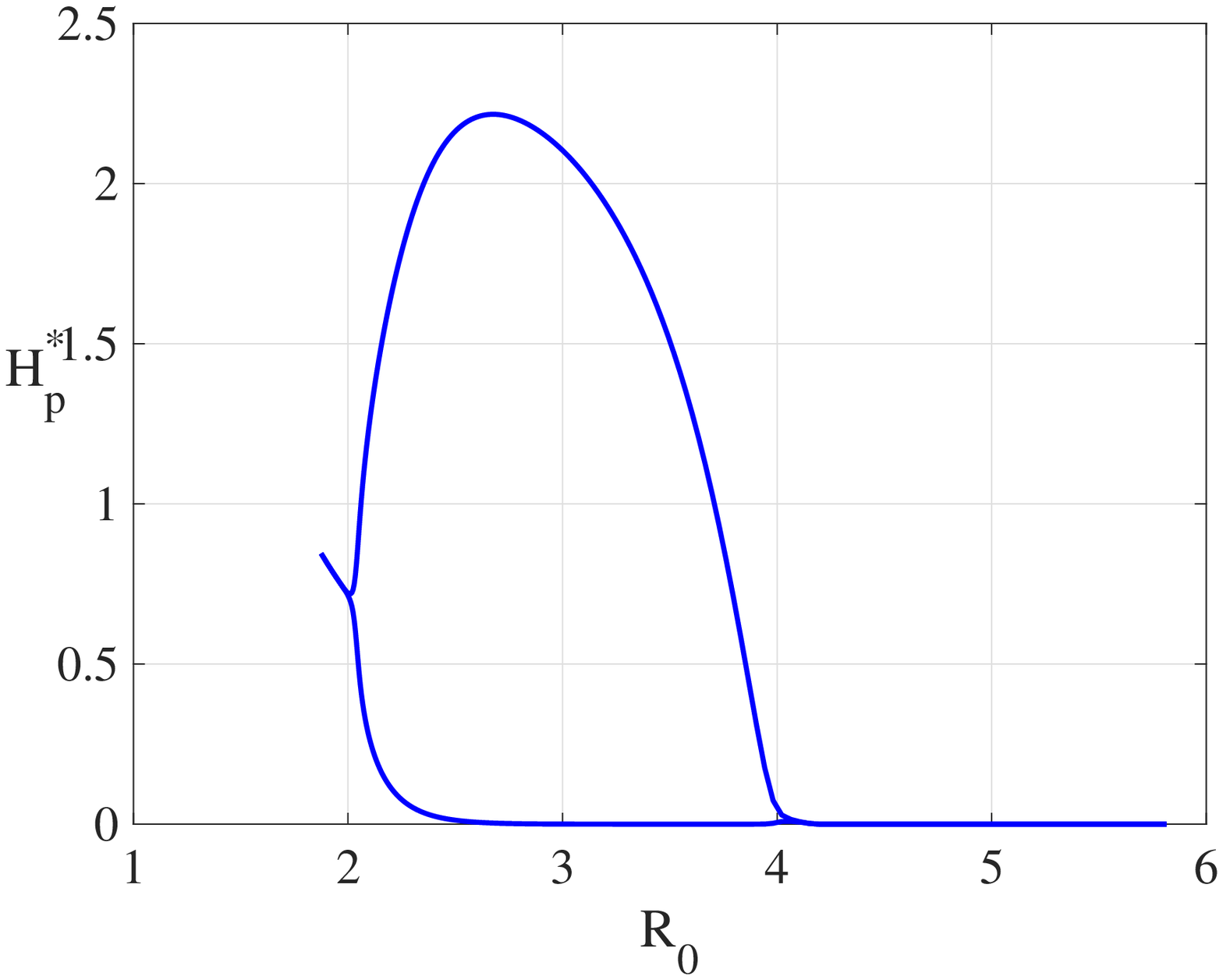}\\
(a) \; Bifurcation with respect to $\gamma$. &  (b) \; Bifurcation with respect to $\Ro$. \\
  \end{tabular}
  \caption{Bifurcation diagram showing the maxima and minima of Healthy plant biomass when $\delta=0.2$}
\label{fig13}
\end{figure}
\begin{figure}[ht]
\begin{tabular}{c@{\hspace{0.5cm}}c}
 \includegraphics[width = 0.5\textwidth]{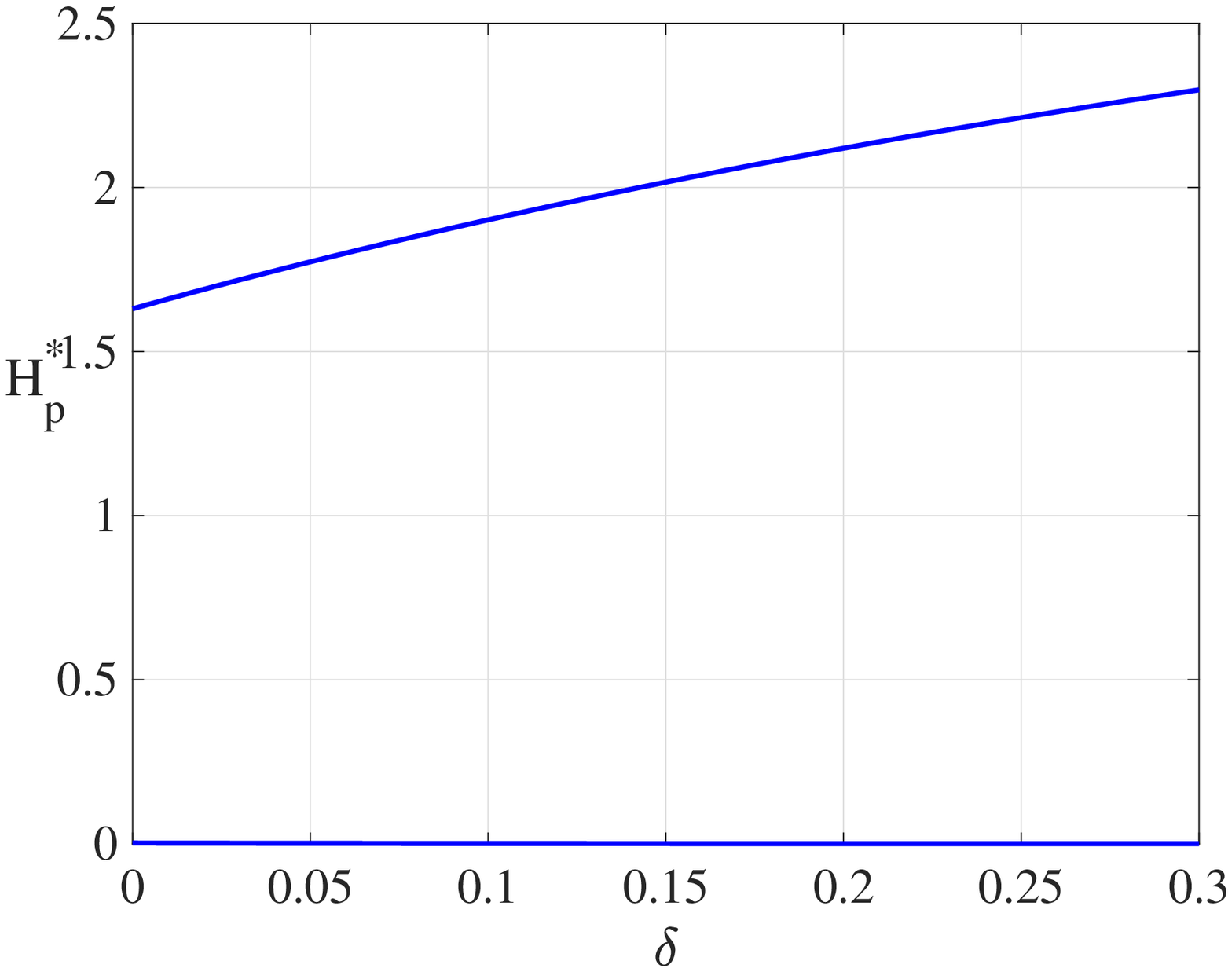}& \includegraphics[width = 0.5\textwidth]{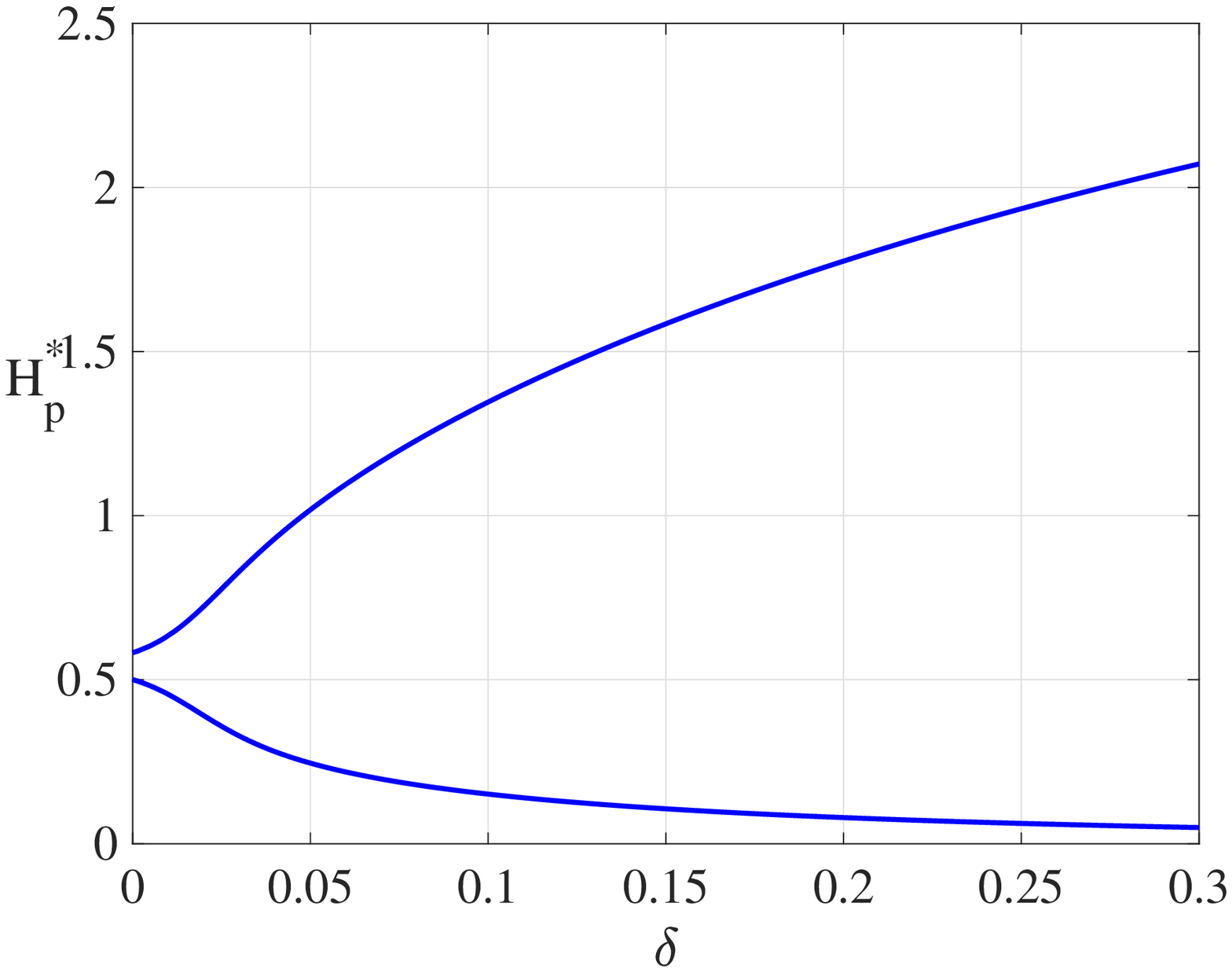}
  \end{tabular}
  \caption{Bifurcation diagram showing the maxima and minima of Healthy plant biomass and Infected vector for different values of $\delta$ with $k_p=0.005$ and (a) $\gamma=0.1$ - (b) $\gamma=0.2$.}
\label{fig10ab}
\end{figure}

In Fig. \ref{fig13}, page \pageref{fig13}, we provide the same simulations with $\delta>0$, namely $\delta=0.2$. Clearly, Hopf bifurcation occurs for smaller and larger values of $\gamma$, compared to the case when $\delta=0$. 
In Fig. \ref{fig10ab}, page \pageref{fig10ab}, we consider the case when $\delta\geq 0$, for a given value of $\gamma$, here $0.1$,  where Hopf bifurcation occurs. Here we need to solve the full system as given in \eqref{model1}. Interestingly, Matlab solvers struggled to solve the full system and adjustments of \textit{RelTol} and \textit{AbsTol} were again necessary.
First, in Fig. \ref{fig10ab}, page \pageref{fig10ab}, with (a), $\gamma=0.1$, and (b), $\gamma=0.2$, we show that the oscillatory behavior is amplified as $\delta$ grows. Thus $\delta>0$ not only increases the interval, for $\gamma$, where Hopf bifurcation can take place, but also amplifies the amplitude of the oscillations.
Last but not least, these numerical results confirm that our approach in the theoretical part, i.e. studying subsystem (\ref{model1alpha0}), page \pageref{model1alpha0}, was appropriate to show that a Hopf bifurcation can occur.

So far, roguing seems to be the best way to control the Cassava Mosaic Disease. However, as our study shows, roguing can amplify a possible periodic behavior in the system. It means that, at some moment, the disease seems to almost disappear, such the decision of stopping the roguing can be taken prematurely. Similar oscillatory behaviour has been observed in plant epidemiological models, see for example \cite{bosch},\cite{holt}. However, it is important to notice that the model presented here is different from these works. In particular, their investigations are based on numerical simulations. 

However, thanks to the sensitivity analysis provided in the previous subsection, some additional control strategies could be taken into account. For instance, reducing $\beta_{vp}$ and $\beta_{pv}$ is possible. In fact, in Kenya, tomato crops are protected against whiteflies using eco-friendly nets \cite{M}. The results are very promising and this control approach seems to be suitable and also sustainable. However, the use of nets is only possible with young cassava plants, less than 1,5m high. The nets also may change locally the environmental parameters (temperature, humidity) of the plants, and thus may have an impact on their (photosynthetic) growth, and eventually, foster other kind of diseases, e.g., fungi. In fact, there exists several other control strategies, including the use of natural enemies, natural pesticides or pesticides extracted from plants, fungi like Beauveria bassiana, etc., and traps. Indeed, traps are widely used around the world, using pheromones or food to attract and catch insects (Anguelov et al., 2017a), and/or for mat- ing disruption (Anguelov et al., 2017b).

Of course, additional bio-control strategies could be considered in our model: for examples, adding resistant plants (and find, like in \cite{Anguelov_PMA_2012}, the right percentage of resistant cassava variety to plant to reduce the impact of the disease), use entomopathogenic fungi  or mycoinsecticides to kill whiteflies, etc. Last but not least, traps, using pheromones, either to catch the insects \cite{Anguelov2017a}, and/or to disrupt mating \cite{Anguelov2017b}. In fact, they are many control strategies that could be tested using modeling and simulations, but also field experiments.

\section{Conclusion}
In this chapter, we presented a modeling example of a crop vector-borne disease with particular focus on cassava and the cassava mosaic virus disease. We showed that interesting mathematical models can be built that require various mathematical tools to be studied. Our model is relatively generic and could be applied to other plant-vector-virus interactions, and, of course, could be improved, taking into account vector aggregation and dispersal, like in \cite{Hebert2016}, or include explicitly the spatial component, like in \cite{chapwanya_dumont}.

\tcr{As alluded earlier, many other control measures may be used, but the proposed model suggests roguing is the most effective way to control the disease. This involves the uprooting of infected plants from the field. However, it has also been reported, see \cite{Legg2015}, that roguing is unpopular among farmers due to the resulting reduction in plant population. Our modeling effort suggests that the induced cycles may be misleading as the farmers may prematurely stop the roguing process. }


\tcm{In addition to roguing, it could be interesting to consider a proportion of a new variety of cassava, less susceptible to the virus or eventually resistant (like in \cite{Anguelov_PMA_2012}), and then estimate theoretically the proportion needed, in a plot to reduce the epidemiological risk.}

However, our interest is not only in the development and the study of new biological control models, but also in the questions highlighted by the modeling process, the analytical results and the numerical simulations. Indeed, despite a comprehensive literature review, various biological processes seem to be partially known, for instance, the transmission rates of the virus from the plant to the whitefly and from the whitefly to the plant. The plant biomass growth rates were only rigorously studied through experiments in Australia. Surprisingly, such rates were not estimated in Africa or South-America despite several projects on the production of several cassava varieties under various conditions. Certainly, these rates may change, in particular, for diseased plants (i.e., estimate for $m_i$), and for plants infested by susceptible whiteflies (i.e., estimate for $\phi$). In contrast, very complex plant models have been calibrated, but only on healthy plants. However, the reader has to be aware that field experiments can be very difficult to conduct, sometimes with uncertainty in the outputs. That is why the use of models (even theoretical ones) can be useful in designing the appropriate experiments in order to feed the models and/or to improve the knowledge on the biological system. Models are not there to replace the field expert, but to support him/her in decision making. 
\tcm{Last but not least, models can also help to test hypothesis about unknown processes and to focus on more specific experiments. However, whatever the models, it is always important to have in mind that they are only approximation of the reality, such that their predictability will always subject to some uncertainty.}

\tcm{We hope that this study on Cassava Mosaic Disease has highlighted some of the challenges to face when dealing with crop (vector-borne) diseases. That is why, to conclude,} we strongly believe that crop diseases modeling can be a new field of research in Mathematical Epidemiology, and, more generally, in Applied Mathematics. 

\vspace{1cm}
\textbf{Acknowledgement}
The authors acknowledge the support of South African DST/NRF SARChI Chair on Mathematical Models and Methods in Bioengineering and Biosciences (M3B2) of the University of Pretoria (South Africa).

The authors thank the anonymous reviewers for their  fruitful comments that greatly improve the initial manuscript.

\newpage

\section{Annexe A}
\label{annexeA}
\subsection{Existence of a full endemic equilibrium for model (\ref{model1})}

To determine the endemic equilibria, we have to solve the following system 
\begin{equation}
\label{system2}
\left\{
\begin{array}{l}
\dfrac{r_{p}H_{p}^{\#}}{1+\phi V^{\#}}-m_{h}H_{p}^{\#}P^{\#}-\beta_{vp}H_{p}^{\#}I_{v}^{\#}-k_{p}H_{p}^{\#}=0,\\
\dfrac{r_{p}I_{p}^{\#}}{1+\phi V^{\#}}+\beta_{vp}H_{p}^{\#}I_{v}^{\#}-m_{i}P^{\#}I_{p}^{\#}-\left(\gamma+k_{p}\right)I_{p}^{\#}=0,\\
\alpha_{v}\left(1+\delta\dfrac{I_{p}^{\#}}{1+I_{p}^{\#}}\right)V^{\#}-\left(\mu_{1}+\mu_{2}V^{\#}\right)S_{v}^{\#}-\beta_{pv}S_{v}^{\#}I_{p}^{\#}=0,\\
\beta_{pv}S_{v}^{\#}I_{p}^{\#}-\left(\mu_{1}+\mu_{2}V^{\#}\right)I_{v}^{\#}=0,
\end{array}
\right.
\end{equation}
where the $^\#$ denote the value at equilibrium. However we can study different cases, like $H_{p}^{\#}>0$ and, in particular, $H_{p}^{\#}=0$, corresponding to the Full Disease Equilibrium, FDE.

Thus, assuming $H_{p}^{\#}=0$ leads to
\[
P^{\#}=I_{p}^{\#}=\dfrac{1}{m_{i}}\left(\dfrac{r}{1+\phi V^{\#}}-\left(\gamma+k_{p}\right)\right),
\]
with 
\begin{equation}
V^{\#}=\dfrac{1}{\mu_{2}}\left(\alpha_{v}\left(1+\delta\dfrac{I_{p}^{\#}}{1+I_{p}^{\#}}\right)-\mu_{1}\right)
\label{eq:Vdieze}
\end{equation}
Thus using (\ref{system2})$_2$, we deduce
\[
m_{i}I_{p}^{\#}+\left(\gamma+k_{p}\right)=\dfrac{}{1+\phi V^{\#}}=\dfrac{r\mu_{2}\left(1+I_{p}^{\#}\right)}{\mu_{2}\left(1+I_{p}^{\#}\right)+\phi\left(\alpha_{v}\left(1+I_{p}^{\#}+\delta I_{p}^{\#}\right)-\mu_{1}\left(1+I_{p}^{\#}\right)\right)}.
\]
Reducing to the same denominator, we derive
\[
\left(m_{i}I_{p}^{\#}+\left(\gamma+k_{p}\right)\right)\left(\left(1+I_{p}^{\#}\right)\left(\mu_{2}+\phi\left(\alpha_{v}-\mu_{1}\right)\right)+\delta\phi\alpha_{v}I_{p}^{\#}\right)=r\mu_{2}\left(1+I_{p}^{\#}\right),
\]
that is 
\[
\left(m_{i}I_{p}^{\#}+\left(\gamma+k_{p}\right)\right)\left(\left(\mu_{2}+\phi\left(\alpha_{v}-\mu_{1}\right)\right)+\left(\mu_{2}+\phi\left(\alpha_{v}-\mu_{1}\right)+\delta\phi\alpha_{v}\right)I_{p}^{\#}\right)=r\mu_{2}\left(1+I_{p}^{\#}\right).
\]
Expanding all terms in the previous equality, we obtain a second order equation
\[
\begin{array}{l}
\left(m_{i}\left(\mu_{2}+\phi\left(\alpha_{v}-\mu_{1}\right)\right)+\left(\gamma+k_{p}\right)\left(\delta\phi\alpha_{v}+\left(\mu_{2}+\phi\left(\alpha_{v}-\mu_{1}\right)\right)\right)-r\mu_{2}\right)I_{p}^{\#}+\\
+m_{i}\left(\left(\mu_{2}+\phi\left(\alpha_{v}-\mu_{1}\right)+\delta\phi\alpha_{v}\right)\right)\left(I_{p}^{\#}\right)^{2}+\left(\left(\gamma+k_{p}\right)\left(\mu_{2}+\phi\alpha_{v}-\mu_{1}\right)-r\mu_{2}\right)=0
\end{array}
\]
That is
\[
\left(B+\delta m_{i}\phi\alpha_{v}\right)\left(I_{p}^{\#}\right)^{2}+\left(B+A+\left(\gamma+k_{p}\right)\phi\alpha_{v}\delta\right)I_{p}^{\#}+A=0,
\]
with
\[
A=\left(\gamma+k_{p}\right)\left(\mu_{2}+\phi\left(\alpha_{v}-\mu_{1}\right)\right)-r\mu_{2}=\phi\left(\gamma+k_{p}\right)\left(\alpha_{v}-\mu_{1}\right)-\left(r-\left(\gamma+k_{p}\right)\right)\mu_{2},
\]
\[
B=m_{i}\left(\mu_{2}+\phi\left(\alpha_{v}-\mu_{1}\right)\right)>0.
\]
Assuming 
\[
\dfrac{\left(\gamma+k_{p}\right)\left(\mu_{2}+\phi\left(\alpha_{v}-\mu_{1}\right)\right)}{r\mu_{2}}<1,
\]
then $A<0$. Then, we compute
\[
\Delta=\left(B+A+\left(\gamma+k_{p}\right)\phi\alpha_{v}\delta\right)^{2}-4A\left(B+\delta m_{i}\phi\alpha_{v}\right)>0,
\]
such that, we deduce the following real positive root
\begin{equation}
I_{p}^{\#}=\dfrac{1}{2\left(B+\delta m_{i}\phi\alpha_{v}\right)}\left(-\left(B+A+\left(\gamma+k_{p}\right)\phi\alpha_{v}\delta\right)+\sqrt{\Delta}\right)>0.
\label{IpFDE}    
\end{equation}
Since we always assume that $r>\gamma+k_{p}$, we can consider the following two cases:
\begin{itemize}
\item when $\phi=0$ (no impact of the vectors on the plant growth rate), FDE always exists. Using (\ref{IpFDE}), we deduce
\[
I_{p}^{\#}=\dfrac{-A}{B}=\dfrac{r_{p}-\left(\gamma+k_{p}\right)}{m_{i}}>0.
\]
Then, the other values $V^{\#}$ (using (\ref{eq:Vdieze})), $S_{v}^{\#}$ (using (\ref{system2}$_{3}$), and $I_{v}^{\#}$ follow. 
\item When $\phi>0$ and such that 
\begin{equation}
\phi<\dfrac{\left(r_{p}-\left(\gamma+k_{p}\right)\right)\mu_{2}}{\left(\gamma+k_{p}\right)\left(\alpha_{v}-\mu_{1}\right)},\label{eq:phi}
\end{equation}
then, according to \eqref{IpFDE}, FDE always exists .
\end{itemize}

\end{document}